\documentclass{cernrep}
\usepackage{graphicx,here} 
\usepackage{amsfonts,amssymb,amsmath}
\font\tenimbf=cmmib10 at 10pt
\font\sevenimbf=cmmib10 at 6pt
\font\fiveimbf=cmmib10 at 4pt
\newfam\imbf
\textfont\imbf=\tenimbf
\scriptfont\imbf=\sevenimbf
\scriptscriptfont\imbf=\fiveimbf

\begin{document}
 
\title{Hard scattering at RHIC: Experimental review
\footnote{$^\star$ Contribution to the CERN Yellow Report on Hard Probes in Heavy Ion Collisions at the LHC}}
\author{David d'Enterria \footnote{$^\dagger$ denterria$@$nevis.columbia.edu}}
\institute{Nevis Laboratories, Columbia University\\ 
Irvington, NY 10533, and New York, NY 10027, USA.}

\maketitle
 
\begin{abstract}
The most significant results on hard scattering processes in Au+Au, p+p, and d+Au
collisions at $\sqrt{s_{NN}}$ = 200 GeV obtained after 3 years of operation 
at the BNL Relativistic Heavy-Ion Collider (RHIC) are summarized. 
%Arguably, the most interesting experimental results of heavy-ion reactions at 
%collider energies are in the where central Au+Au data show a very different behaviour compared to p+p, d+Au, 
%and peripheral Au+Au collisions.
Hadron production at high transverse momentum ($p_{T}$) in central Au+Au collisions
shows very different properties (production yields, flavor composition, azimuthal correlations, ...) 
in comparison to p+p, d+Au, and peripheral Au+Au collisions,
and this fact constitutes arguably the most interesting outcome of the experimental 
program so far. Four main topics are covered in this report:
(i) the depletion of high $p_{T}$ inclusive hadron spectra, %(jet leading particles), 
(ii) the ``anomalous'' baryon-meson composition at intermediate $p_{T}$'s, 
(iii) the disappearance of away-side azimuthal (jet) correlations, and
(iv) the large value of collective azimuthal (elliptic flow) correlations.
Additionally, the observed high $p_{T}$ ``Cronin enhancement'' in d+Au collisions
will be discussed. A succinct comparison of experimental data to theoretical models 
is also provided.
\end{abstract}

\section{Introduction}

Heavy-ion collisions at RHIC collider energies aim at the study
of QCD matter at extreme energy densities. The driving force of this research
program is the study of the transition from hadronic matter to 
a deconfined, chirally symmetric plasma of quarks and gluons (QGP) 
predicted to occur above a critical energy density $\epsilon\approx$ 0.7 $\pm$ 0.3 GeV/fm$^3$ 
by lattice QCD calculations~\cite{latt}. Additionally, the combination 
of high-energies and large nuclear systems in the initial-state provides 
propitious grounds for the study of the regime of non-linear many-body parton 
(mostly gluon) dynamics at small-$x$ where saturation effects are expected to 
play a dominant role %(see Section 4.x and 
~\cite{cgc}. The most interesting 
phenomena observed so far in Au+Au reactions at $\sqrt{s_{_{NN}}}$ = 130 and 
200 GeV are in the high transverse momentum ($p_{T} \gtrsim$ 2 GeV/$c$) sector 
where the production of hadrons in central collisions shows substantial 
differences compared to more elementary reactions either in the ``vacuum'' 
(p+p, $e^+e^-$) or in a cold nuclear matter environment (d+Au).

Several arguments make of hard scattering processes an excellent experimental 
probe in heavy-ion collisions:

\begin{itemize}
\item They occur in the first instants of the reaction process ($\tau\sim 1/p_{T}\lesssim$ 0.1 fm/$c$) 
in parton-parton scatterings with large $Q^2$ and, thus, constitute direct probes 
of the partonic phase(s) of the reaction.
\item A direct comparison to the baseline ``vacuum'' (p+p) production yields 
is straightforward after scaling by the nuclear geometry (in the simplest approach, 
this is done scaling by the nuclear overlap function $T_{AB}$ or, equivalently, by 
the number of nucleon-nucleon ($NN$) binary inelastic collisions, 
$N_{coll} = T_{AB}\cdot \sigma_{NN}$, given by a Glauber model).
\item The production yields of high $p_{T}$ particles are theoretically calculable 
via perturbative (or classical-field) QCD methods.
\end{itemize}

Thus, high $p_{T}$ particles constitute experimentally and theoretically well calibrated 
observables that are sensitive to the properties of the dense QCD medium existing in 
central heavy-ion collisions.

\newpage

The main observations so far at RHIC are the following:

\begin{itemize}
\item  The high $p_{T}$ yields of inclusive charged hadrons and $\pi^0$ in central Au+Au 
at $\sqrt{s_{_{NN}}}$ = 130~\cite{phenix_hipt_130,star_hipt_130,phenix_hipt_130_2} 
and 200 GeV~\cite{phenix_pi0_200,star_hipt_200,phenix_hipt_200,phobos_hipt_200,brahms_hipt_200},
are suppressed by as much as a factor 4 -- 5 compared to p+p and peripheral Au+Au yields
scaled by $T_{AB}$ (or $N_{coll}$).

\item At intermediate $p_{T}$'s ($p_{T}\approx$ 2. -- 4. GeV/$c$) in central Au+Au, 
at variance with mesons ($\pi^0$~\cite{phenix_pi0_200} and kaons~\cite{star_hipt_strange_200})
no suppression is seen for baryons ($p,\bar{p}$~\cite{phenix_ppbar_130,phenix_ppbar_200} 
and $\Lambda,\bar{\Lambda}$~\cite{star_hipt_strange_200}), yielding 
an ``anomalous'' baryon over meson ratio $p/\pi\sim$ 1 much larger than the ``perturbative'' 
$p/\pi\sim$ 0.1 -- 0.3 ratio observed in p+p collisions~\cite{ISR_ppbar,ISR_pi0} and in $e^+e^-$ 
jet fragmentation~\cite{DELPHI}. 

\item The near-side azimuthal correlations of high $p_{T}$ (leading) hadrons 
emitted in central and peripheral Au+Au reactions~\cite{star_away_side,chiu_qm02} 
are, on the one hand, clearly reminiscent of jet-like parton fragmentation as found in p+p 
collisions. On the other, away-side azimuthal correlations (from back-to-back jets) in central 
Au+Au collisions are found to be significantly suppressed %compared to p+p 
\cite{star_away_side}.

\item At low $p_{T}$ the strength of the azimuthal anisotropy parameter $v_{2}$ is found to
be large and consistent with hydrodynamical expectations for elliptic flow. 
Above $p_{T}\sim$ 2 GeV/$c$ where the contribution from collective behaviour is 
negligible, $v_{2}$ has still a sizeable value with a flat (or slightly decreasing) 
behaviour as a function of $p_{T}$~\cite{star_hipt_strange_200,star_hipt_flow_130,phenix_flow}.

\item High $p_{T}$ production in ``cold nuclear matter'' as probed in 
d+Au reactions~\cite{phenix_dAu,star_dAu,phobos_dAu,brahms_hipt_200} 
not only is not suppressed %(at variance with Au+Au central)
but it is {\it enhanced} compared to p+p collisions, in a way very much reminiscent of the
``Cronin enhancement'' observed in p+A collisions at lower center-of-mass energies~\cite{cronin}.
\end{itemize}

All these results point to strong medium effects at work in central 
Au+Au collisions, and have triggered extensive theoretical discussions
based on perturbative %(see Sections X,Y of this Report) 
or ``classical''-field QCD. Most of the studies on the high $p_{T}$ suppression are 
based on the prediction~\cite{Gyu90} that a deconfined and dense medium would induce 
multiple gluon radiation off the scattered partons, effectively leading to a depletion 
of the high-$p_{T}$ hadronic fragmentation products (``jet quenching''), %see Sections 4.3 of this Report), 
though alternative interpretations have been also put forward 
based on initial-state gluon saturation effects (``Color Glass Condensate'', CGC) ~\cite{cgc,dima}, %(see Section 4.42 and
or final-state hadronic reinteractions~\cite{gallmeister}. 
The different behaviour of baryons and mesons at moderately high $p_{T}$'s 
has been interpreted, among others, in terms of ``quark recombination'' (or coalescence)
effects in a thermalized partonic (QGP-like) medium %(see Section 4.41 of this Report and
~\cite{recomb}, whereas the disappearance of the back-to-back azimuthal correlations 
can be explained in both QGP energy loss and CGC monojet scenarios. Finally, the large
value of $v_{2}$ above 2 GeV/$c$ has been addressed by jet energy loss~\cite{glv_flow}, gluon
saturation~\cite{cgc_flow}, and quark recombination~\cite{recomb_flow} models.

This summary report presents the $p_{T}$, $\sqrt{s_{NN}}$, centrality, particle-species,
and rapidity dependence of the inclusive high $p_{T}$ particle production, plus the
characteristics of the produced jets and collective elliptic flow signals
extracted from the azimuthal correlations at large $p_{T}$, 
as measured by the four experiments at RHIC (BRAHMS, PHENIX, PHOBOS and STAR) 
in Au+Au, p+p, and d+Au collisions. The whole set of experimental data puts
strong constraints on the different proposed physical explanations for 
the underlying QCD medium produced in heavy-ion collisions at RHIC and at LHC energies.\\

%%%%%%%%%%%%%%%%%%%%%%%%%%%%%%%%%%%%%%%%%%%%%%

\section{High $p_{T}$ hadron production in p+p collisions at $\sqrt{s}$ = 200 GeV}

\subsection{p+p inclusive cross-sections:}

Proton-proton collisions are the baseline ``vacuum'' reference to which one
compares the Au+Au results in order to extract information about the QCD medium properties.
At $\sqrt{s}$ = 200 GeV, there currently exist three published measurements 
of high $p_{T}$ hadron cross-sections in $p+p(\bar{p})$ collisions: UA1 $p+\bar{p}\rightarrow h^\pm$ 
($|\eta|<2.5$, $p_{T} <$ 7 GeV/$c$)~\cite{UA1}, PHENIX $p+p\rightarrow \pi^0$ ($|\eta|<0.35$, $p_{T}<$ 14 GeV/$c$)
~\cite{phenix_pp_pi0_200}, and STAR $p+p\rightarrow  h^\pm$ ($|\eta|<0.5$, $p_{T}<$ 10 GeV/$c$)
~\cite{star_hipt_200}. At $\sqrt{s}$ = 130 GeV, an interpolation between the ISR
inclusive charged hadron cross-section and UA1 and FERMILAB data, has been also used as a
reference for Au+Au at this value of $\sqrt{s}$. Globally the spectra can be 
reasonably well parametrized by a power-law form  $A \cdot (1+p_{T} /p_0)^{-n}$ with 
the parameters\footnote{The fit parameters $p_0$ and $n$ are actually strongly correlated 
via the mean $p_{T}$ of the collision: $\langle p_{T} \rangle = 2p_{0}/(n-3)$.} 
reported in Table \ref{tab:powerlaw_fits}.

\begin{table}[htb]
\begin{center}
\begin{tabular}{l|c|c|c|c|c}
\hline\hline
\hspace{1mm} 
system & $\sqrt{s}$ (GeV) &  $p_{T}^{min}$ (GeV/$c$) & $A$ (mb GeV$^{-2}c^{3}$) &  $p_0$ (GeV/$c$) & $n$ \\\hline
 $p+p \rightarrow h^\pm$ (\small{inel., interpolation}~\cite{phenix_hipt_130}) &  130  & 0.4 & 330 & 1.72  &  12.40\\
 $p+\bar{p} \rightarrow h^\pm$ (NSD, UA1)~\cite{UA1}			&  200  & 0.25 	& 286 	& 1.80  &  12.14\\
 $p+p \rightarrow h^\pm$ (NSD, STAR)~\cite{star_hipt_200}		&  200  & 0.4 	& 286 	& 1.43 	&  10.35\\
 $p+p \rightarrow \pi^0$ (inel., PHENIX)~\cite{phenix_pp_pi0_200} 	&  200  & 1.0 	& 386 	& 1.22  &   9.99\\
\hline\hline
\end{tabular}
\label{tab:powerlaw_fits}
\end{center}
\vspace*{-0.6cm}
\caption{Parameters of the fit $ Ed^{3}\sigma/dp^{3} = A \cdot (1+p_{T} /p_0)^{-n}$ to the inclusive 
$p_{T}$ distributions of all existing $p+p(\bar{p})$ hadron (inelastic or non-singly diffractive) 
cross-sections measurements at $\sqrt{s}$ = 200 GeV.}
\end{table}

\begin{figure}[htbp]
\begin{center}
\begin{minipage}[t]{75mm}
\includegraphics[height=9.cm]{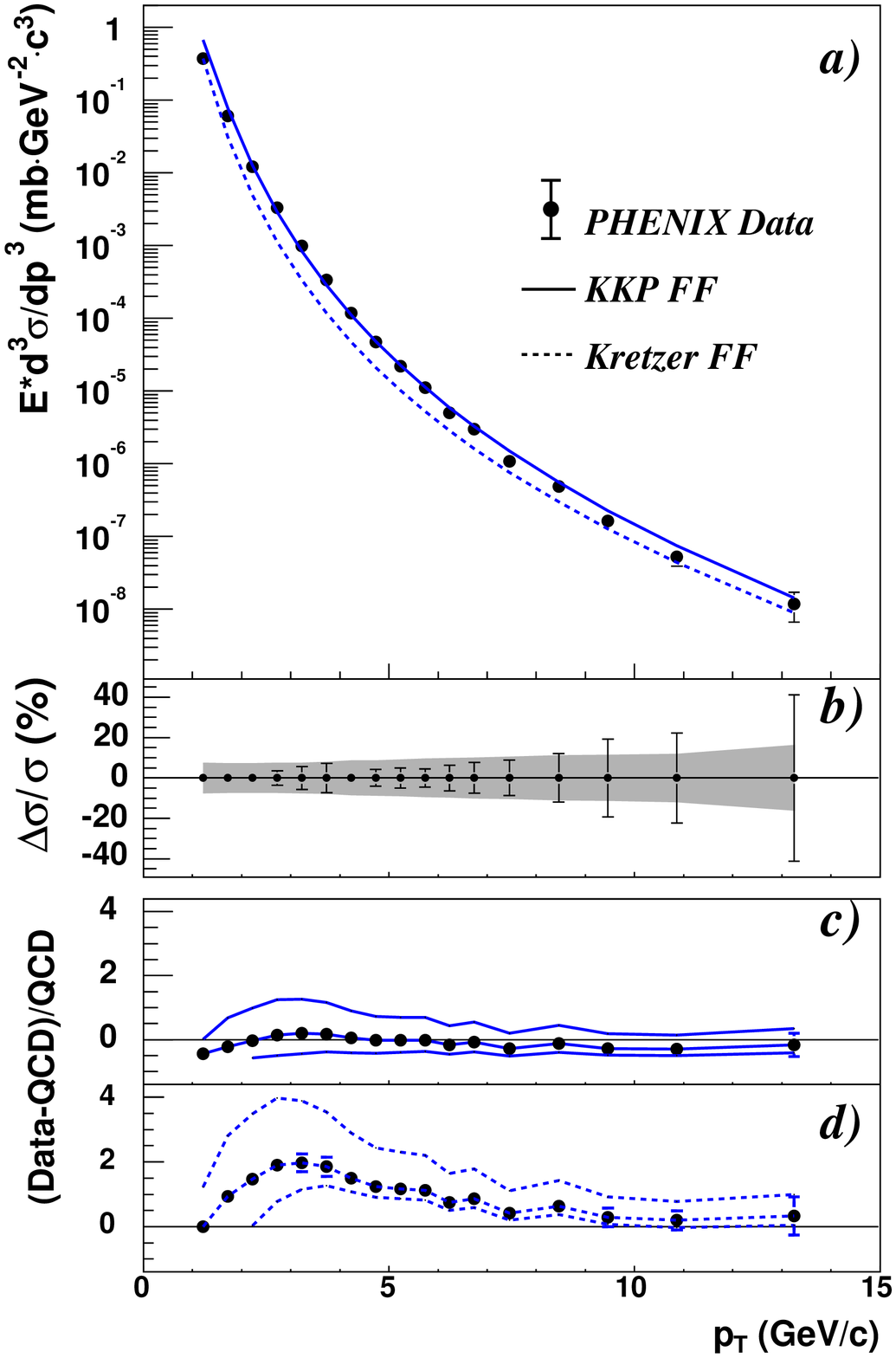}
\end{minipage}
\begin{minipage}[t]{75mm}
\includegraphics[height=9.6cm]{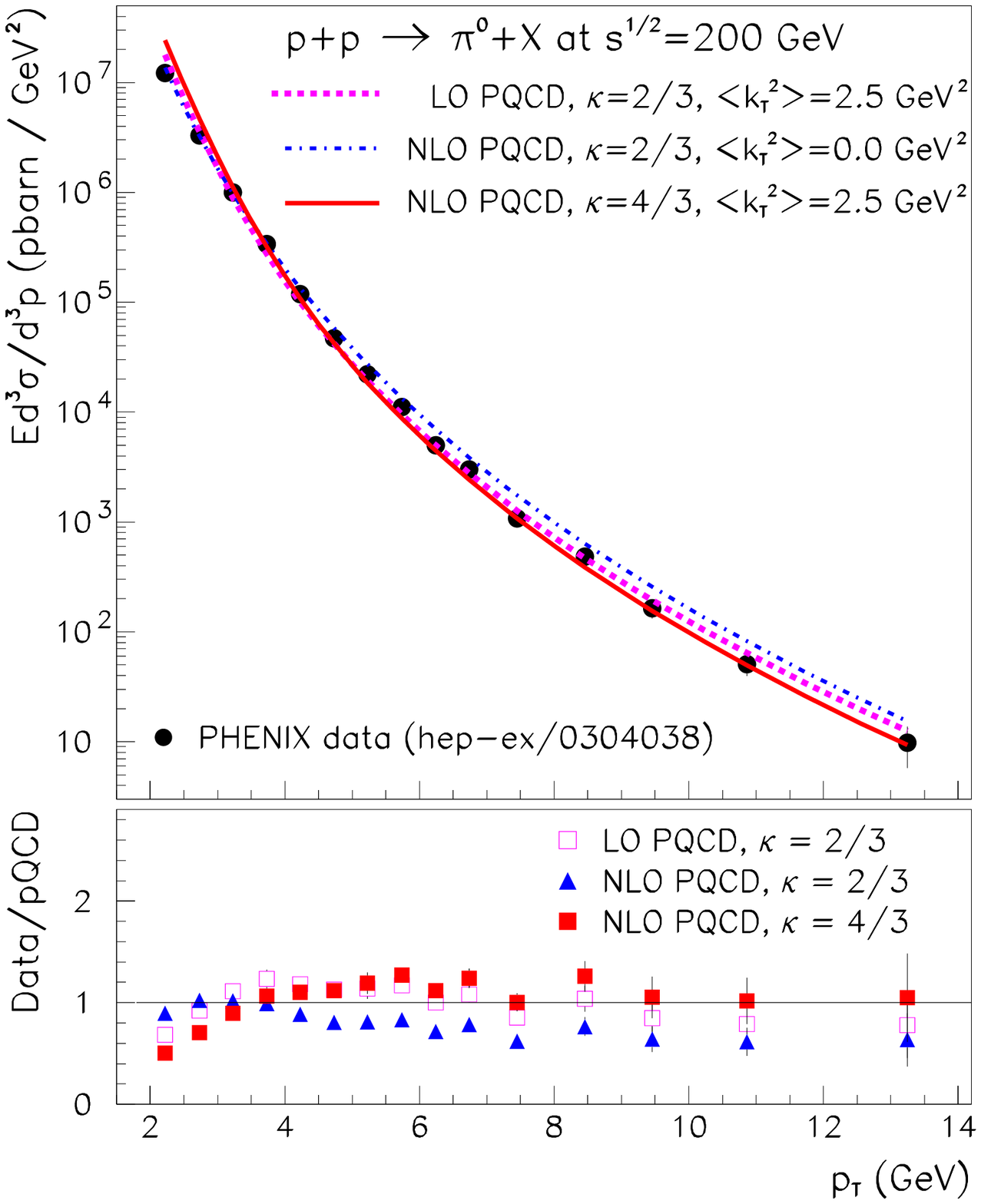}%star_chargedhad_spectra_pp.eps}
\end{minipage}
%\hspace*{-0.8cm}
\end{center}
\vspace*{-0.5cm}
\caption{High $p_{T}$ $\pi^0$ cross-section in p+p collisions at $\sqrt{s}$ = 200 GeV
(PHENIX) compared to the results of two different NLO pQCD calculations: 
\cite{phenix_pp_pi0_200} ({\it left}),~\cite{levai_pp_pi0} ({\it right}).}
\label{fig:pp_pi0_vs_pQCD}
\end{figure}

In general, all experimental results are consistent within each other, although it is 
claimed~\cite{star_hipt_200} that STAR p+p inclusive charged yield is smaller by a factor of 
0.79 $\pm$ 0.18 compared to UA1 $p+\bar{p}$ results (approximately independent of $p_{T}$), 
the difference due in large part to differing non-singly-diffractive\footnote{PHENIX 
high $p_{T}$ $\pi^0$ cross-section is inclusive and contains, in principle, all 
inelastic (including diffractive) channels.} (NSD) cross section measured 
(35 $\pm$ 1 mb~\cite{UA1} in the first and 30.0 $\pm$ 3.5 mb~\cite{star_hipt_200} in the later). 
Standard next-to-leading-order (NLO) perturbative QCD calculations %(see Section 2.21 of this Report) 
describe well the available high $p_{T}$ p+p data at $\sqrt{s}$ = 200 GeV 
(see Fig. \ref{fig:pp_pi0_vs_pQCD} for $\pi^0$).% The quality of the data permits to 

\vspace*{-0.5cm}

\subsection{p+p azimuthal correlations:}

PHENIX~\cite{rak} has studied the azimuthal correlations at high $p_{T}$ in p+p
collisions at $\sqrt{s}$ = 200 GeV extracting several parameters characterizing 
the produced jets:

\begin{itemize}
\item Mean jet fragmentation transverse momentum: $\langle |j_{\perp y}|\rangle$ 
= 373 $\pm$ 16 MeV/$c$, %. This value is 
in agreement with previous measurements at ISR~\cite{ccor_jperp}
and showing no significant trend with increasing $\sqrt{s}$.
\item Average parton transverse momentum (fitted to a constant above 1.5 GeV/$c$): 
$\langle |k_{\perp y}|\rangle$ = 725 $\pm$ 34 MeV/$c$. The momentum of the pair 
$p_{\perp}$ is related to the individual parton $\langle |k_{\perp y}|\rangle$ via 
$\sqrt{\langle |p_{\perp}^{2}|\rangle_{pair}}\, = \, \sqrt{2\pi} \,\langle |k_{\perp y}|\rangle$.
The extracted $\sqrt{\langle |p_{\perp}^{2}|\rangle_{pair}}$ = 1.82 $\pm$ 0.85 GeV/$c$ is
in agreement with the existing systematics of dimuon, diphoton and dijet data in 
hadronic collisions~\cite{apana99}.
\end{itemize}

%%%%%%%%%%%%%%%%%%%%%%%%%%%%%%%%%%%%%%%%%%%%%%

\section{High $p_{T}$ hadron production yields in Au+Au collisions}

There is a significant amount of high $p_{T}$ Au+Au experimental spectra ($p_{T}>$ 2 GeV/$c$) measured 
by the 4 experiments at RHIC: inclusive charged hadrons at 130~\cite{phenix_hipt_130,star_hipt_130,phenix_hipt_130_2} 
and 200 GeV~\cite{star_hipt_200,phenix_hipt_200,phobos_hipt_200,brahms_hipt_200}, neutral pions
at 130~\cite{phenix_hipt_130} and 200 GeV~\cite{phenix_pi0_200}, protons and
antiprotons at 130~\cite{phenix_ppbar_130} and 200 GeV~\cite{phenix_ppbar_200},
$K^0_s$ at 200 GeV~\cite{star_hipt_strange_200}, and
$\Lambda,\bar{\Lambda}$ at 200 GeV~\cite{star_hipt_strange_200}. Moreover, all these 
spectra are measured for different centrality bins and permit to address
the impact parameter dependence of high $p_{T}$ production.

Details on hadron production mechanisms in $AA$ are usually studied
via their scaling behavior with respect to p+p collisions. On the one hand, 
soft processes ($p_{T}<$ 1 GeV/$c$) are expected to 
scale with the number of participating nucleons $N_{part}$~\cite{wnm}
(and they actually approximately do~\cite{phenix_dNdy,phobos_dNdy}). 
On the other, in the framework of collinear factorization, hard processes are 
incoherent and thus expected to scale with $N_{coll}$. 

\begin{figure}[htbp]
\begin{center}
\begin{minipage}[t]{75mm}
\includegraphics[height=6.0cm]{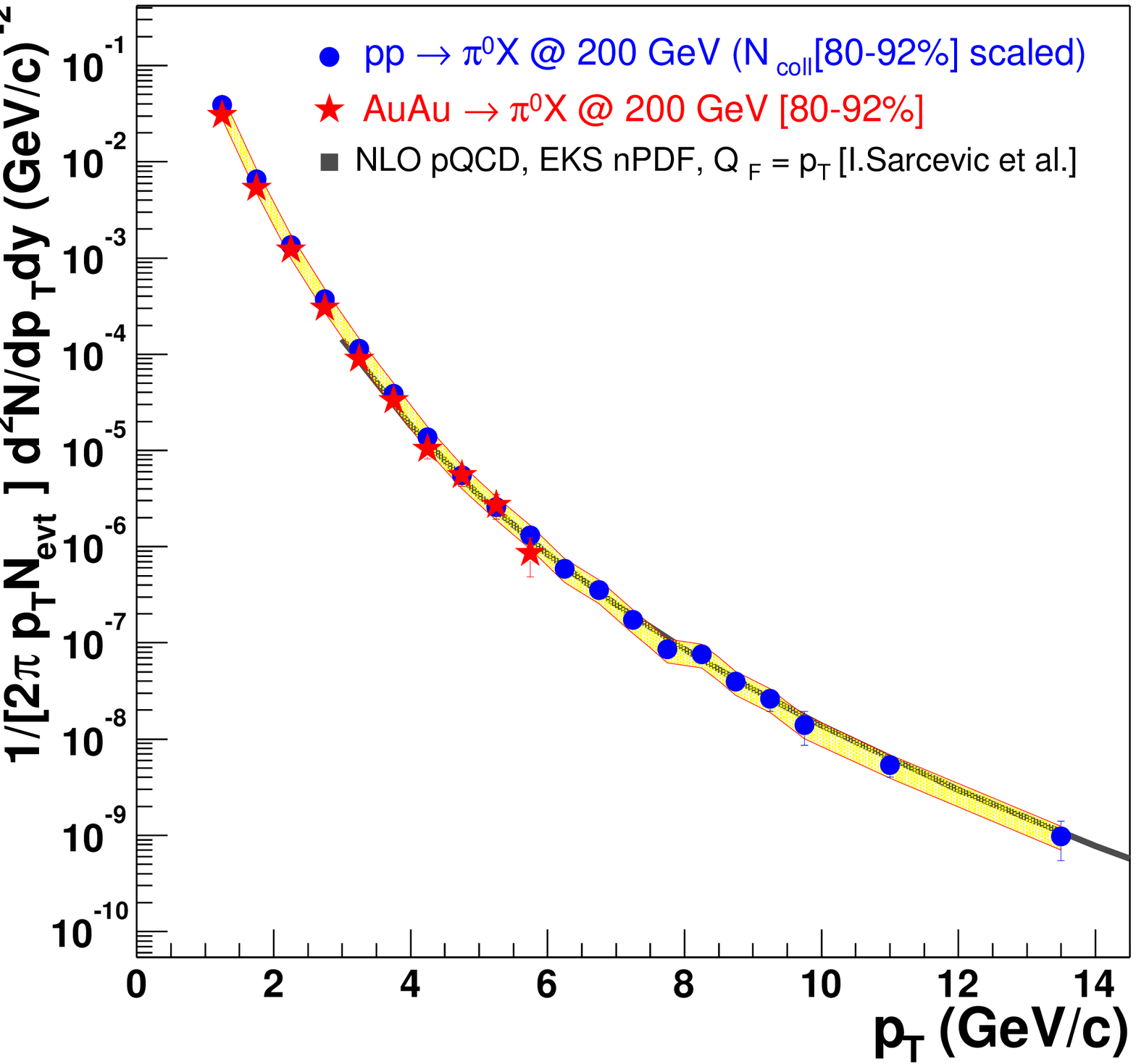}
\end{minipage}
\begin{minipage}[t]{75mm}
\includegraphics[height=6.0cm]{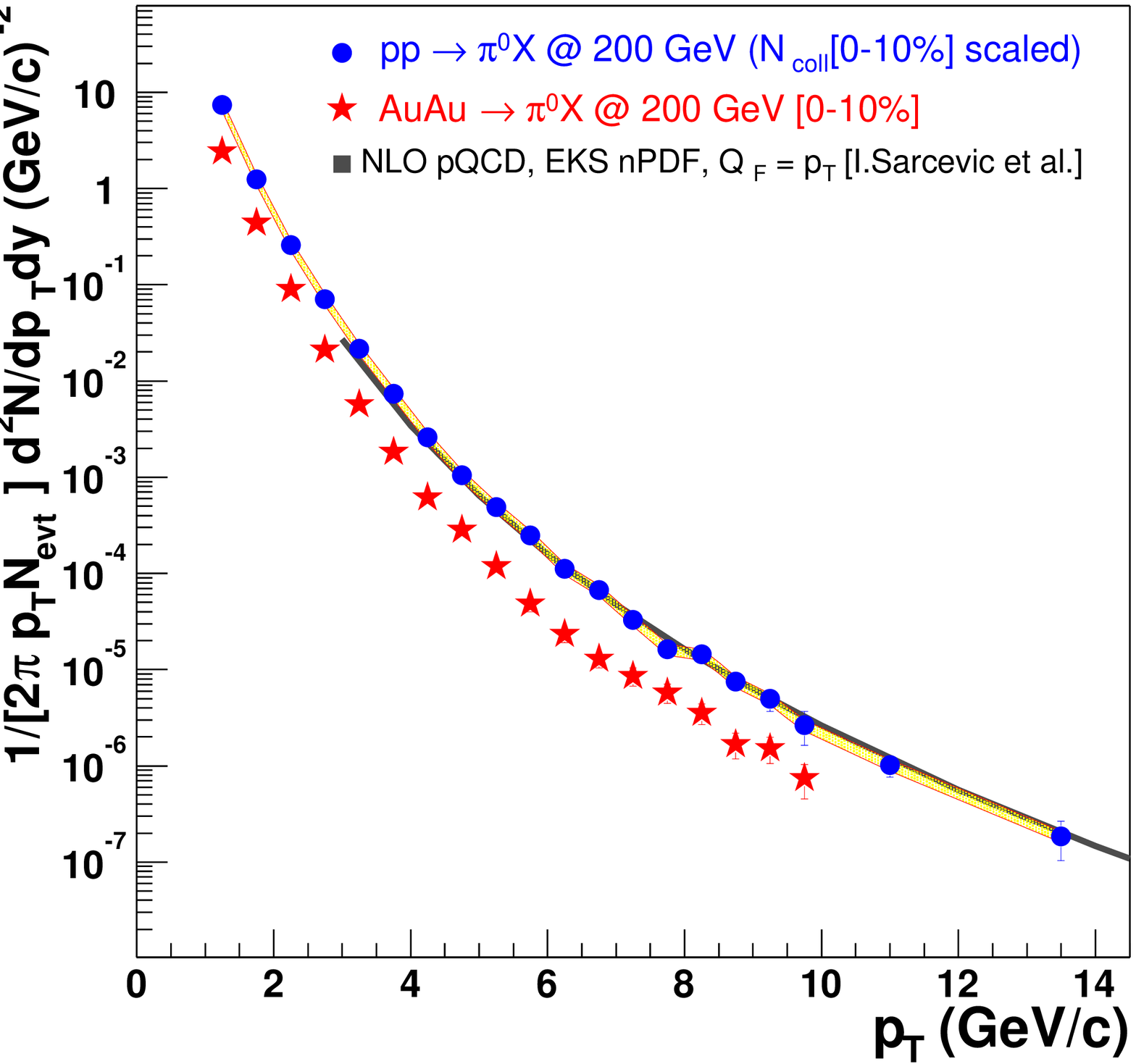}
\end{minipage}
\end{center}
\vspace*{-0.3cm}
\caption{Invariant $\pi^0$ yields measured by PHENIX in peripheral ({\it left}) and in central ({\it right}) 
Au+Au collisions (stars), compared to the $N_{coll}$ scaled p+p $\pi^0$ yields (circles) and to
a NLO pQCD calculation~\cite{ina} (gray line). The yellow band around the scaled p+p points 
includes in quadrature the absolute normalization errors 
in the p+p and Au+Au spectra as well as the uncertainties in $T_{AB}$. 
Updated version of Fig. 1 of \protect\cite{dde_qm02} with final published data 
\protect\cite{phenix_pi0_200,phenix_pp_pi0_200}.}
\label{fig:phenix_pi0_pp_AuAu}
\end{figure}

The first interesting result at RHIC in the high $p_{T}$ 
sector is the {\it breakdown} of this $N_{coll}$ scaling for central Au+Au collisions. 
Fig. \ref{fig:phenix_pi0_pp_AuAu} shows the comparison of the p+p $\pi^0$ spectrum 
to peripheral (left) and central (right) Au+Au spectra, and to pQCD calculations.
Whereas peripheral data is consistent with a simple superposition of individual
$NN$ collisions, central data shows a suppression factor of 4 -- 5 with respect
to this expectation.

It is customary to quantify the medium effects at high $p_{T}$ using the {\it nuclear modification factor} 
given by the ratio of the $AA$ to the p+p invariant yields scaled by the nuclear geometry ($T_{AB}$):
\begin{equation} 
R_{AA}(p_{T})\,=\,\frac{d^2N^{\pi^0}_{AA}/dy dp_{T}}{\langle T_{AB}\rangle\,\times\, d^2\sigma^{\pi^0}_{pp}/dy dp_{T}}.
\label{eq:R_AA}
\end{equation}
$R_{AA}(p_{T})$ measures the deviation of $AA$ from an incoherent superposition 
of $NN$ collisions in terms of suppression ($R_{AA}<$1) or enhancement ($R_{AA}>$1).
%The next subsections summarize the current knowledge of $R_{AA}$

%%%%%%%%%%%%%%%%%%%%%%%%%%%%%%%%%%%%%%%%%%%%%%

\subsection{High $p_{T}$ suppression: magnitude and $p_{T}$ dependence}

\begin{figure}[htbp]%[H]
\begin{minipage}[t]{75mm}
\includegraphics[height=7.0cm]{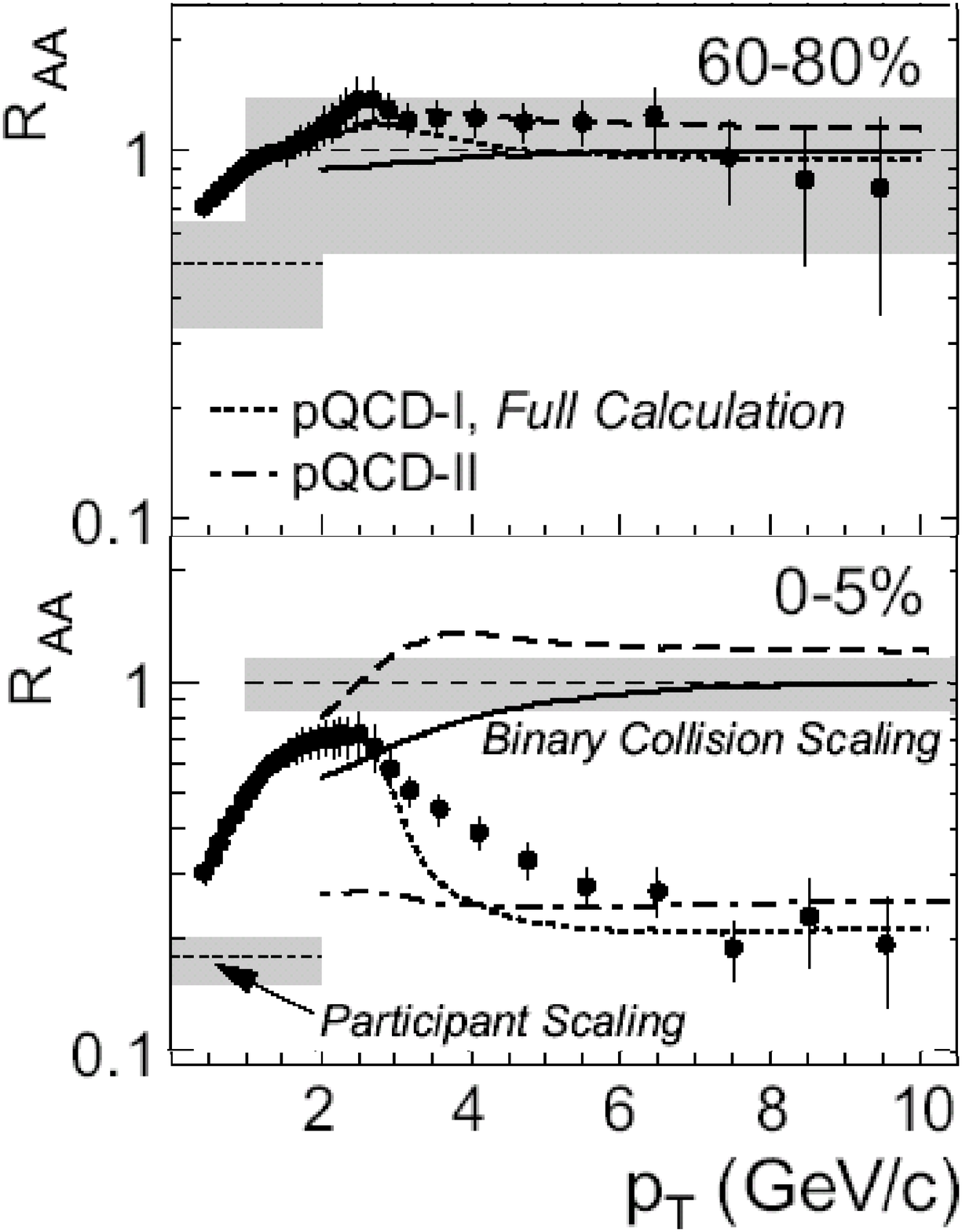}
\end{minipage}
\begin{minipage}[t]{75mm}
\includegraphics[height=6.8cm]{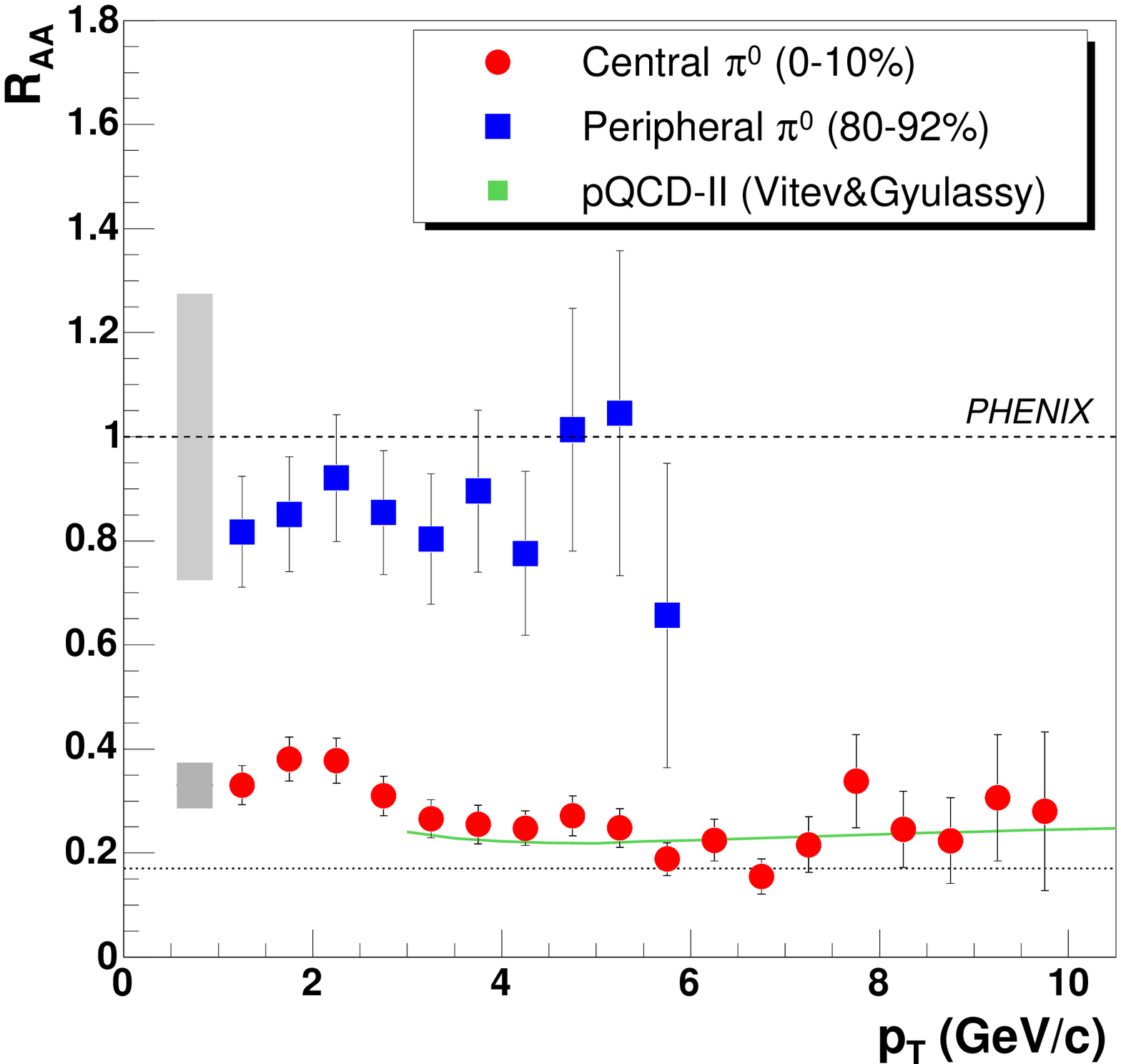}
\end{minipage}
\vspace*{-0.3cm}
\caption[]{Nuclear modification factor, $R_{AA}(p_{T})$, in peripheral and central Au+Au reactions
for charged hadrons ({\it left}) and $\pi^0$ ({\it right}) measured at $\sqrt{s_{NN}}$ = 200 GeV
by STAR and PHENIX respectively. A comparison to theoretical curves: pQCD-I~\cite{vitev}, 
pQCD-II~\cite{xnwang}, is also shown.}
\label{fig:R_AA_phenix_star}
\end{figure}

Figure~\ref{fig:R_AA_phenix_star} shows $R_{AA}(p_{T})$ for 
$h^\pm$ (STAR~\cite{star_hipt_200}, left) and $\pi^0$ (PHENIX~\cite{phenix_pi0_200}, right)
measured in peripheral (upper points) and central (lower points) Au+Au reactions at 
$\sqrt{s_{NN}}$ = 200 GeV. As seen in Fig. \ref{fig:phenix_pi0_pp_AuAu}, peripheral collisions are 
consistent\footnote{Although peripheral STAR charged hadron data seems 
to be slightly above $R_{AA}$ = 1 and PHENIX $\pi^0$ data seems to be below, within errors 
both measurements are consistent with ``collision scaling''.} with p+p collisions plus 
$N_{coll}$ scaling as well as with standard pQCD calculations~\cite{vitev,xnwang}, while 
central Au+Au are clearly suppressed by a factor $\sim$ 4 -- 5.

The high $p_{T}$ suppression in central collisions for both $\pi^0$ and $h^\pm$ is smallest at 
$p_{T}$ = 2 GeV/$c$ and increases to an approximately constant suppression factor of 
1/$R_{AA}\approx$~4 -- 5 over $p_{T}$~=~5 -- 10~GeV/$c$. {\it Above} 5 GeV/$c$ 
the data are consistent within in errors with ``participant scaling'' given by
the dotted line at $R_{AA}\approx$ 0.17 in both plots (actually both STAR and PHENIX data are 
systematically slightly above this scaling). The magnitude and $p_{T}$ dependence of $R_{AA}$ in 
the range $p_{T}$ = 1 -- 10 GeV/$c$ (corresponding to parton fractional momenta 
$x_{1,2}=p_{T}/\sqrt{s}(e^{\pm y_{1}}+e^{\pm y_{2}})\approx 2p_{T}/\sqrt{s}\sim$ 
0.02 -- 0.1 at midrapidity), is alone inconsistent with ``conventional'' nuclear effects like 
leading-twist shadowing of the nuclear parton distribution functions (PDFs)~\cite{eks98,vogt}.
Different pQCD-based jet quenching calculations~\cite{vitev,xnwang,arleo,levai_pp_pi0,urs} based on 
medium-induced radiative energy loss, can reproduce the {\it magnitude} of the $\pi^0$ 
suppression assuming the formation of a hot and dense partonic system characterized by 
different, but related, properties: i) large initial gluon densities $dN^{g}/dy\sim$ 1000
~\cite{vitev}, ii) large ``transport coefficients'' $\hat{q}_{0}\sim$ 3.5  GeV/fm$^2$~\cite{arleo}, 
iii) high opacities $L/\lambda\sim$ 3 -- 4~\cite{levai_pp_pi0}, or iv) effective parton 
energy losses of the order of $dE/dx\sim$ 14 GeV/fm~\cite{xnwang}. 

The {\it $p_{T}$ dependence} of the quenching predicted by all models that include
the QCD version of the Landau-Pomeranchuck-Migdal (LPM) interference effect (BDMPS~\cite{BDMPS} 
and GLV~\cite{glv} approaches) is a slowly (logarithmic) increasing function of $p_{T}$, 
a trend not compatible with the data over the entire measured $p_{T}$ range. Other approaches, such 
as constant energy loss per parton scattering, are also not supported as discussed in~\cite{ina}. 
Analyses which combine LPM jet quenching together with shadowing and initial-state $p_{T}$ broadening 
(``pQCD-II''~\cite{vitev} in Fig. \ref{fig:R_AA_phenix_star}) globally 
reproduce the observed flat $p_{T}$ dependence of $R_{AA}$, as do recent approaches 
that take into account detailed balance between parton emission and absorption 
(``pQCD-I''~\cite{xnwang} in Fig. \ref{fig:R_AA_phenix_star}, left). 

At variance with parton energy loss descriptions, a gluon saturation calculation~\cite{dima}
is able to predict the magnitude of the observed suppression, although it fails to reproduce 
exactly the flat $p_{T}$ dependence of the quenching~\cite{star_hipt_200}. Similarly,
{\it semi-quantitative} estimates of final-state interactions in a dense {\it hadronic} 
medium~\cite{gallmeister} yield the same amount of quenching as models based on partonic 
energy loss, however it is not yet clear whether the $p_{T}$ evolution of the hadronic 
quenching factor is consistent with the data or not~\cite{star_hipt_200,xnwang03}.

The amount of suppression for $\pi^0$~\cite{phenix_pi0_200} and $h^\pm$~\cite{star_hipt_200,phenix_hipt_200} 
is the same above $p_{T}\approx$ 4 -- 5 GeV/$c$ for all centrality classes~\cite{phenix_hipt_200}. 
However, below $p_{T}\sim$ 5 GeV/$c$, $\pi^0$'s are more suppressed than inclusive charged hadrons 
in central collisions (as can be seen by comparing the right and left plots of Fig. 
\ref{fig:R_AA_phenix_star}). This is due to the enhanced baryon production contributing 
to the total charged hadron yield in the intermediate $p_{T}$ region 
($p_{T}\approx$ 1 -- 4 GeV/$c$) in Au+Au collisions~\cite{phenix_ppbar_200,star_hipt_strange_200} 
(see section \ref{hadron_composition} below).

%\begin{figure}[htbp]%[H]
%\includegraphics[height=7.0cm]{figs/RAA_pi0_vs_hadr_phenix.eps}
%\caption[]{$R_{AA}$ for $(h^{+} + h^{-})/2$ and $\pi^{0}$ as function of $p_{T}$
%for minimum bias and 9 centrality classes (PHENIX)~\protect\cite{phenix_hipt}.}
%\label{fig:R_AA_phenix_star}
%\end{figure}

%%%%%%%%%%%%%%%%%%%%%%%%%%%%%%%%%%%%%%%%%%%%%%
%\clearpage

\subsection{High $p_{T}$ suppression: $\sqrt{s_{NN}}$ dependence}

\begin{figure}[htbp]%[H]
%\vspace*{-.8cm}
\begin{center}
\includegraphics[height=6.2cm]{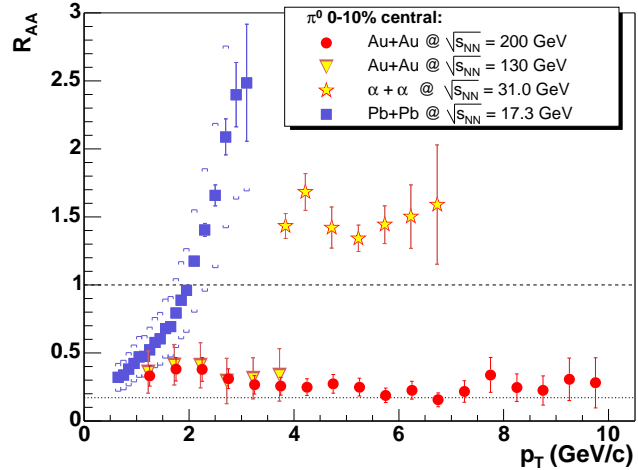}
\end{center}
\vspace*{-.5cm}
\caption[]{Nuclear modification factor, $R_{AA}(p_{T})$, for $\pi^0$ measured in
central ion-ion reactions at CERN-SPS~\protect\cite{wa98_pi0} (squares), CERN-ISR~\protect\cite{ISR_pi0} (stars), 
and BNL-RHIC (triangles~\protect\cite{phenix_hipt_130}, circles~\protect\cite{phenix_pi0_200}) energies.}
%: (i) CERN-SPS central Pb+Pb~\cite{wa98_pi0} (squares), (ii) CERN-ISR $\alpha+\alpha$~\cite{isr_pi0}
%(stars), (iii) PHENIX-RHIC central Au+Au at 130 GeV (crosses), and (iv) PHENIX-RHIC central 
%Au+Au at 200 GeV (circles).}
\label{fig:R_AA_pi0_syst}
\end{figure}

Figure~\ref{fig:R_AA_pi0_syst} shows $R_{AA}(p_{T})$ for several $\pi^0$ measurements 
in high-energy $AA$ collisions at different center-of-mass energies~\cite{dde_breckenridge03}. 
The PHENIX $R_{AA}(p_{T})$ values for central Au+Au collisions at 200 GeV (circles) 
and 130 GeV (triangles) are noticeably below unity in contrast to the enhanced production 
($R_{AA}>$1) observed at CERN-ISR (min. bias $\alpha+\alpha$~\cite{ISR_pi0}, stars) and 
CERN-SPS energies (central Pb+Pb~\protect\cite{wa98_pi0}, squares) and interpreted in terms 
of initial-state $p_{T}$ broadening (``Cronin effect''~\cite{cronin}).

\begin{figure}[htbp]
\begin{center}
\includegraphics[height=6.0cm]{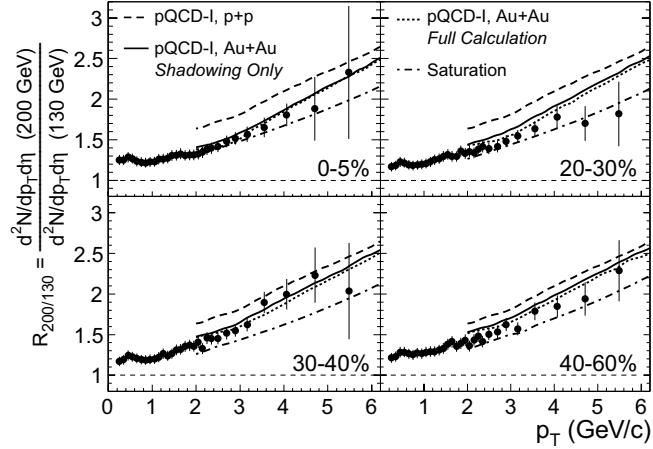}
\end{center}
\vspace*{-.8cm}
\caption{$R_{200/130}(p_{T})$ vs. $p_{T}$ for $(h^++h^-)/2$ for four different centrality bins
measured by STAR compared to pQCD and gluon saturation model predictions~\protect\cite{star_hipt_200}.}
\label{fig:star_hipt_200_130}
\end{figure}

Figure \ref{fig:star_hipt_200_130} shows $R_{200/130}(p_{T})$, the ratio of Au+Au charged hadron 
yields at $\sqrt{s_\mathrm{_{NN}}}$ = 130 and 200 GeV in 4 centrality classes, compared to
pQCD and gluon saturation model predictions~\cite{star_hipt_200}. The increase in high $p_{T}$ 
yields between the two center-of-mass energies is a factor $\sim$2 at the highest $p_{T}$'s, whereas 
at low $p_{T}$, the increase is much moderate, of the order of 15\%. The large increment of the hard 
cross sections is naturally consistent with pQCD expectations due to the increased jet
contributions at high transverse momenta. In the saturation model~\cite{dima} the increase
at high $p_{T}$ is accounted for by the enhanced gluon densities at $\sqrt{s_\mathrm{_{NN}}}$ = 
200 GeV compared to 130 GeV in the ``anomalous dimension'' $x_{T}$ region of the Au 
parton distribution function.

\begin{figure}[htbp]
\begin{center}
\includegraphics[height=5.8cm]{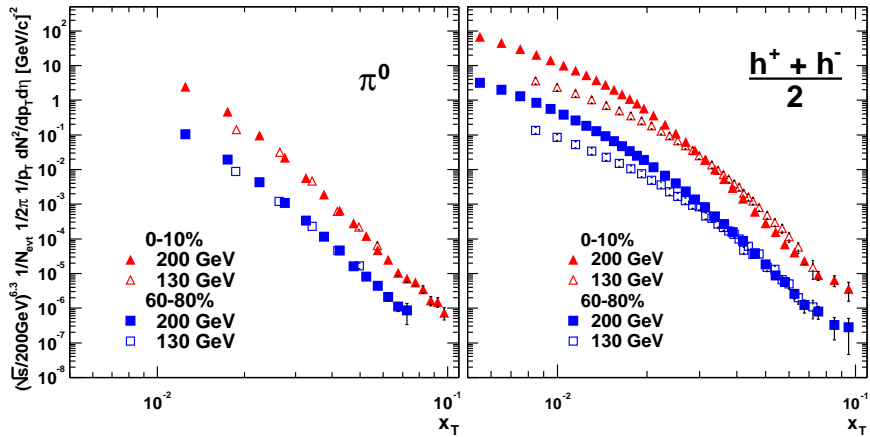}
\end{center}
\vspace*{-0.7cm}
\caption{$x_{T}$ scaled spectra for $\pi^0$ ({\it left}) and $(h^+ + h^-)/2$ ({\it right})
measured in central and peripheral collisions at $\sqrt{s_{_{NN}}}$ = 130 and 200 GeV by PHENIX~\cite{phenix_hipt_200}.
Central (Peripheral) $x_{T}$ spectra are represented by triangles (squares), and
solid (open) symbols represent $x_{T}$ spectra from $\sqrt{s_{_{NN}}}$ = 200 GeV 
($\sqrt{s_{_{NN}}}$ = 130 GeV scaled by a factor of $[130/200]^{6.3}$).}
\label{fig:xT_scaling_phenix}
\end{figure}

PHENIX~\cite{phenix_hipt_200} has addressed the $\sqrt{s}$ dependence of high $p_{T}$
production by testing the validity of ``$x_{T}$ scaling'' in Au+Au, i.e. verifying the
parton model prediction that hard scattering cross sections can be factorized in 2 terms
depending on $\sqrt{s}$ and $x_{T} = 2p_{T}/\sqrt{s}$ respectively: 
\begin{equation}
 E \frac{d^3\sigma}{dp^3}=\frac{1}{p_{T}^{n(\sqrt{s})}} \: F(x_{T}) \,\,
 \Longrightarrow E \frac{d^3\sigma}{dp^3}={1\over {\sqrt{s}^{{\,n(x_{T},\sqrt{s})}} }} \: G({x_{T}})\,.
\label{eq:x_{T}}
\end{equation}
In (\ref{eq:x_{T}}), $F(x_{T})$ embodies all the $x_{T}$ dependence coming from the parton distribution (PDF) 
and fragmentation (FF) functions\footnote{PDFs and FFs, to first order, scale as the ratio of $p_{T}$ at 
different $\sqrt{s}$.}, while the exponent $n$, related to the underlying parton-parton
scattering, is measured to be $n\approx$ 4 -- 8 in a wide range of $p+p,\bar{p}$ collisions~\cite{phenix_hipt_200}.
% Since $x_{T} = p_{T}/(2\sqrt{s})$, one finds  $E\,d\sigma/d^3p = (1/\sqrt{s}^{n(x_{T},\sqrt{s})}) G(x_{T})$.
Fig.~\ref{fig:xT_scaling_phenix} compares the $x_{T}$-scaled hadron yields in $\sqrt{s_{NN}}$ = 130 GeV 
and 200 GeV Au+Au central and peripheral collisions. According to Eq. (\ref{eq:x_{T}}), 
the ratio of inclusive cross-sections at fixed $x_{T}$ should equal $(200/130)^{n}$.
On the one hand, $x_{T}$ scaling holds\footnote{In the kinematical region, $x_{T}>0.03$, 
where pQCD is expected to hold.} in Au+Au with the same scaling power $n= 6.3\pm0.6$ 
for neutral pions (in central and peripheral collisions) and charged hadrons 
(in peripheral collisions) as measured in p+p~\cite{phenix_hipt_200}. 
This is consistent with equal (pQCD-like) production dynamics 
in p+p and Au+Au, and disfavours final-state effects described with medium-modified
FF's that violate $x_{T}$ scaling (e.g. constant parton energy losses independent of the 
parton $p_{T}$). Equivalently, models that predict strong initial-state effects
(e.g. gluon saturation) respect $x_{T}$ scaling as long as their predicted modified 
nuclear PDFs are depleted, independently of $\sqrt{s}$, by the same amount at a 
given $x_{T}$ (and centrality). On the other hand, Fig.~\ref{fig:xT_scaling_phenix} 
(right) shows that charged hadrons in central collisions (triangles) break $x_{T}$ scaling
which is indicative of a non perturbative modification of particle composition
spectra from that of p+p at intermediate $p_{T}$'s (see section \ref{hadron_composition} below).

%%%%%%%%%%%%%%%%%%%%%%%%%%%%%%%%%%%%%%%%%%%%%%
\clearpage

\subsection{High $p_{T}$ suppression: centrality dependence}

\begin{figure}[htbp]
\begin{center}
%\vspace*{-0.5cm}
%\hspace*{-1.cm}
\begin{minipage}[t]{75mm}
\includegraphics[height=5.5cm]{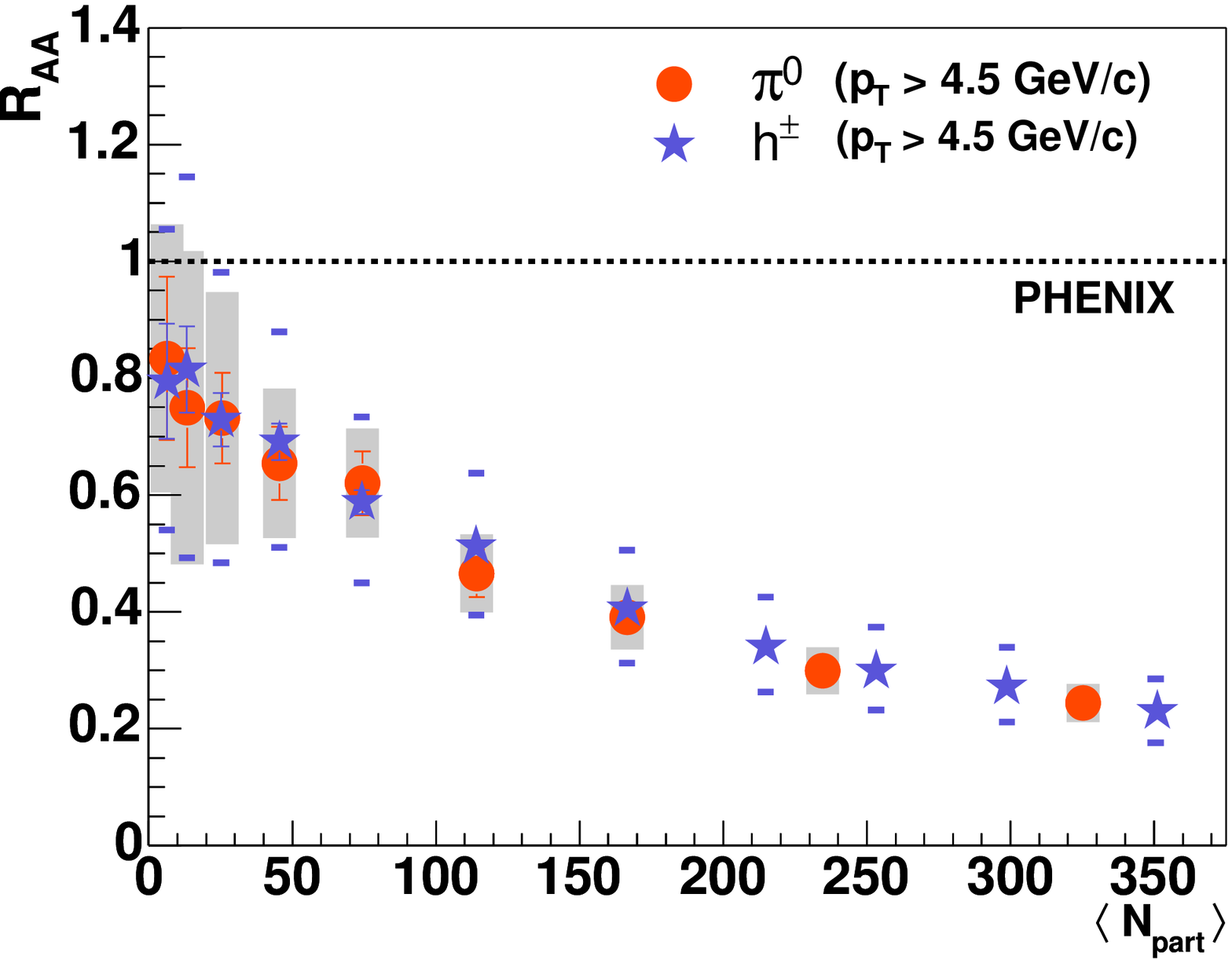}
\end{minipage}
\hspace*{.1cm}
\begin{minipage}[t]{75mm}
\includegraphics[height=5.3cm]{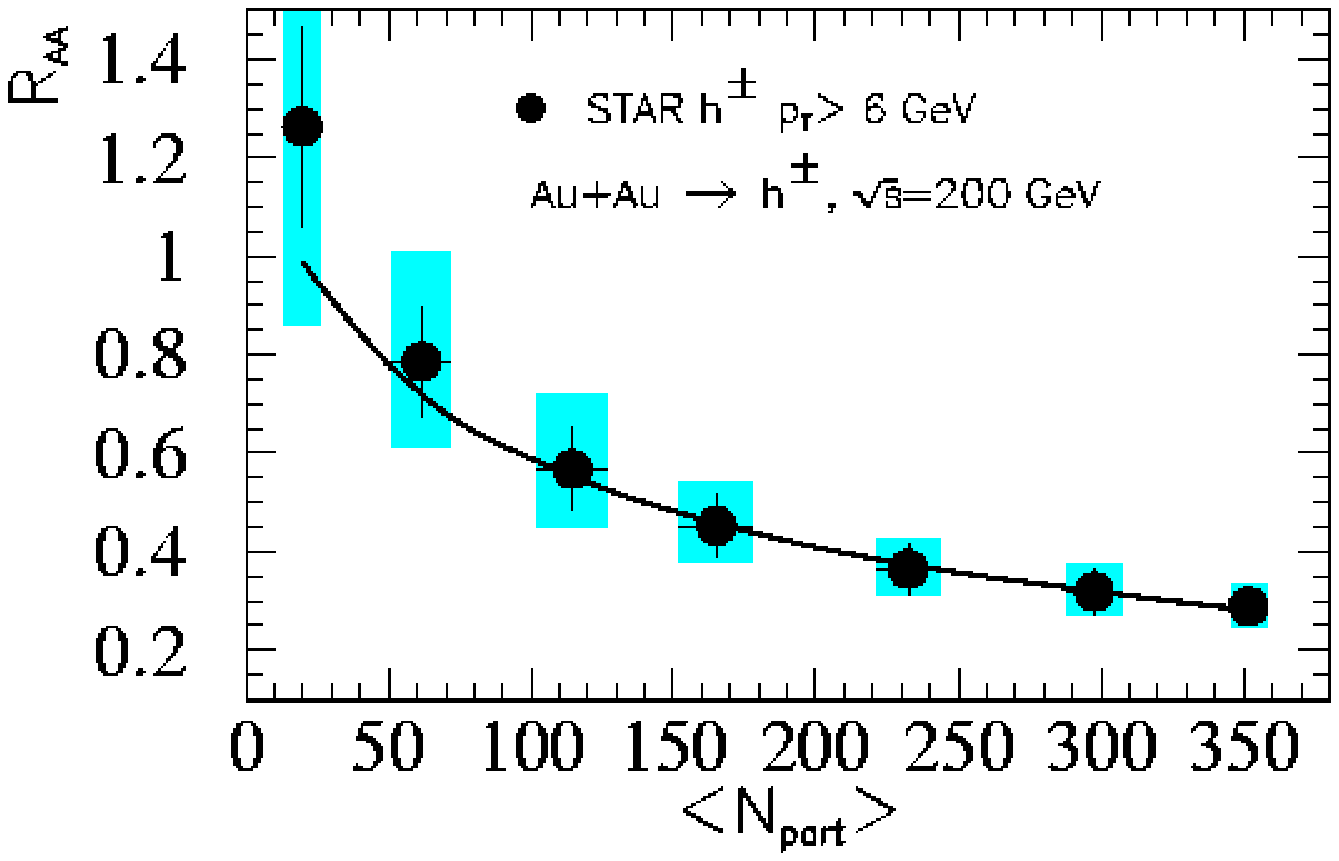}
\end{minipage}
\end{center}
\vspace*{-0.7cm}
\caption[]{Evolution of the high $p_{T}$ $\pi^0$ and $h^\pm$ suppression, $R_{AA}(p_{T}>$ 4.5 GeV/$c$), 
as a function of centrality given by $\langle N_{part} \rangle$ (PHENIX, {\it left}).
Same evolution shown as $R_{AA}(p_{T}>$6.0 GeV/$c$) for STAR $h^\pm$ data~\cite{xnwang03} ({\it right}).}
\label{fig:R_AA_vs_cent}
\end{figure}

In each centrality bin, the value of the high $p_{T}$ suppression can be
quantified by the ratio of Au+Au over $N_{coll}$-scaled p+p yields integrated above 
a given (large enough) $p_{T}$. The centrality dependence of the high $p_{T}$ 
suppression for $\pi^0$ and charged hadrons, given by $R_{AA}(p_{T}>$ 4.5 GeV/$c$), 
is shown in Fig.~\ref{fig:R_AA_vs_cent} (left) as a function of $\langle N_{part} \rangle$ for
PHENIX data. The transition from the $N_{coll}$ scaling behaviour ($R_{AA}\sim$~1) apparent in the 
most peripheral region, $\langle N_{part} \rangle \lesssim$ 40, to the strong suppression 
seen in central reactions ($R_{AA}\sim$~0.2) is smooth. Whether there is 
an abrupt or gradual departure from $N_{coll}$ scaling in the peripheral range cannot be 
ascertained within the present experimental uncertainties~\cite{dde_breckenridge03}. 
The data, however, is inconsistent with $N_{coll}$ scaling (at a $2\sigma$ level) for 
the 40--60\% centrality corresponding to $\langle N_{part} \rangle\approx$ 40 -- 80
~\cite{phenix_hipt_130_2,dde_breckenridge03}, whose estimated ``Bjorken'' energy 
density ($\epsilon_{Bj}\approx$ 1 GeV/fm$^3$)~\cite{dde_breckenridge03} is
in the ball-park of %theoretical estimates for %: (i) the minimal amount of nucleons  
%needed to obtain parton ``percolation'' at RHIC energies ($N_{part}\sim$ 80)~\cite{satz}, and(ii) 
the expected ``critical'' QCD energy density. A similar centrality dependence
of the high $p_{T}$ suppression is seen in STAR $h^\pm$ data (Fig.~\ref{fig:R_AA_vs_cent}, right)

\begin{figure}[htbp]
\hspace*{-2.8cm}
\begin{center}
\begin{minipage}[t]{75mm}
\includegraphics[height=5.6cm]{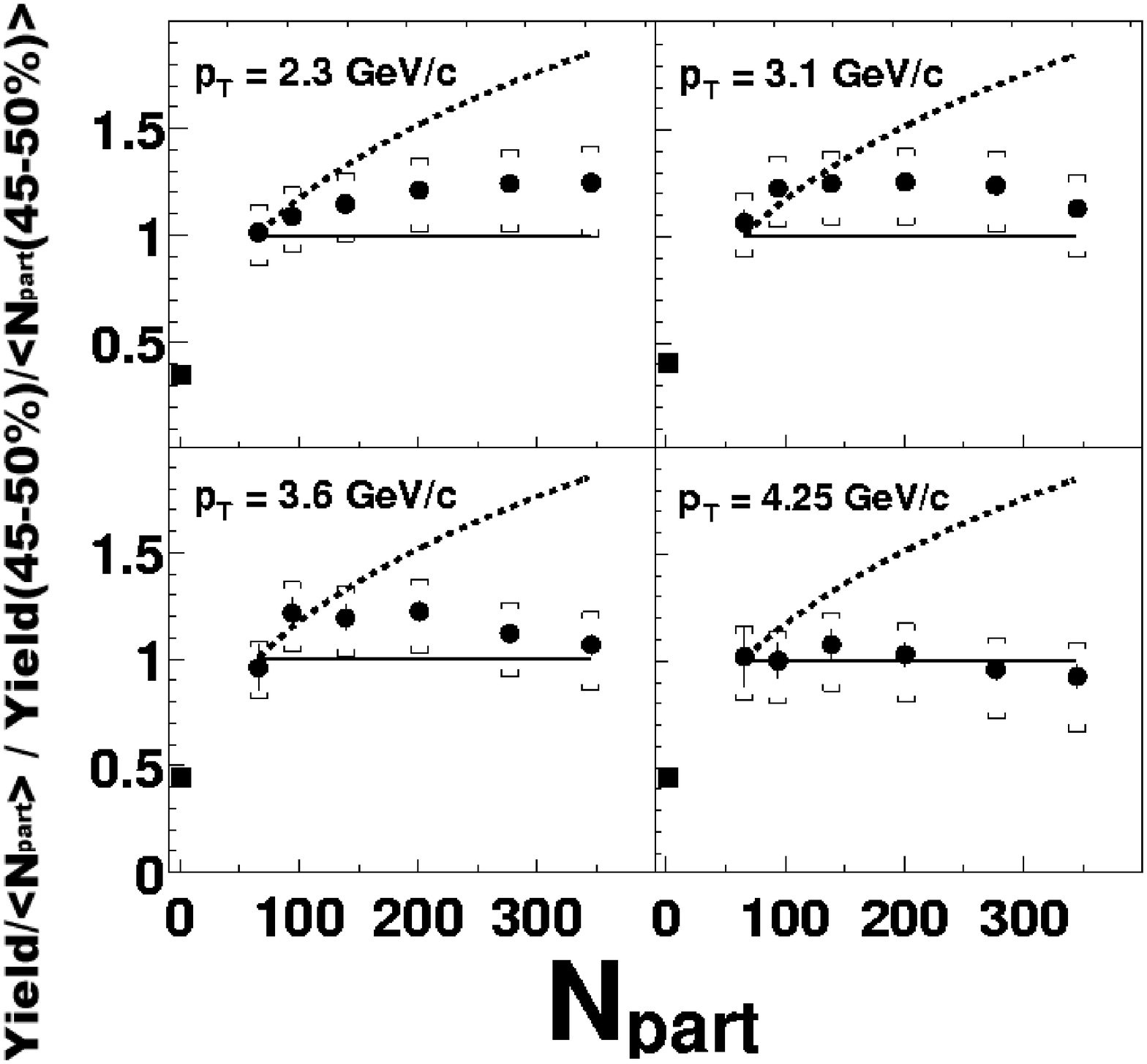}
\end{minipage}
\hspace*{.3cm}
\begin{minipage}[t]{75mm}
\includegraphics[height=5.4cm]{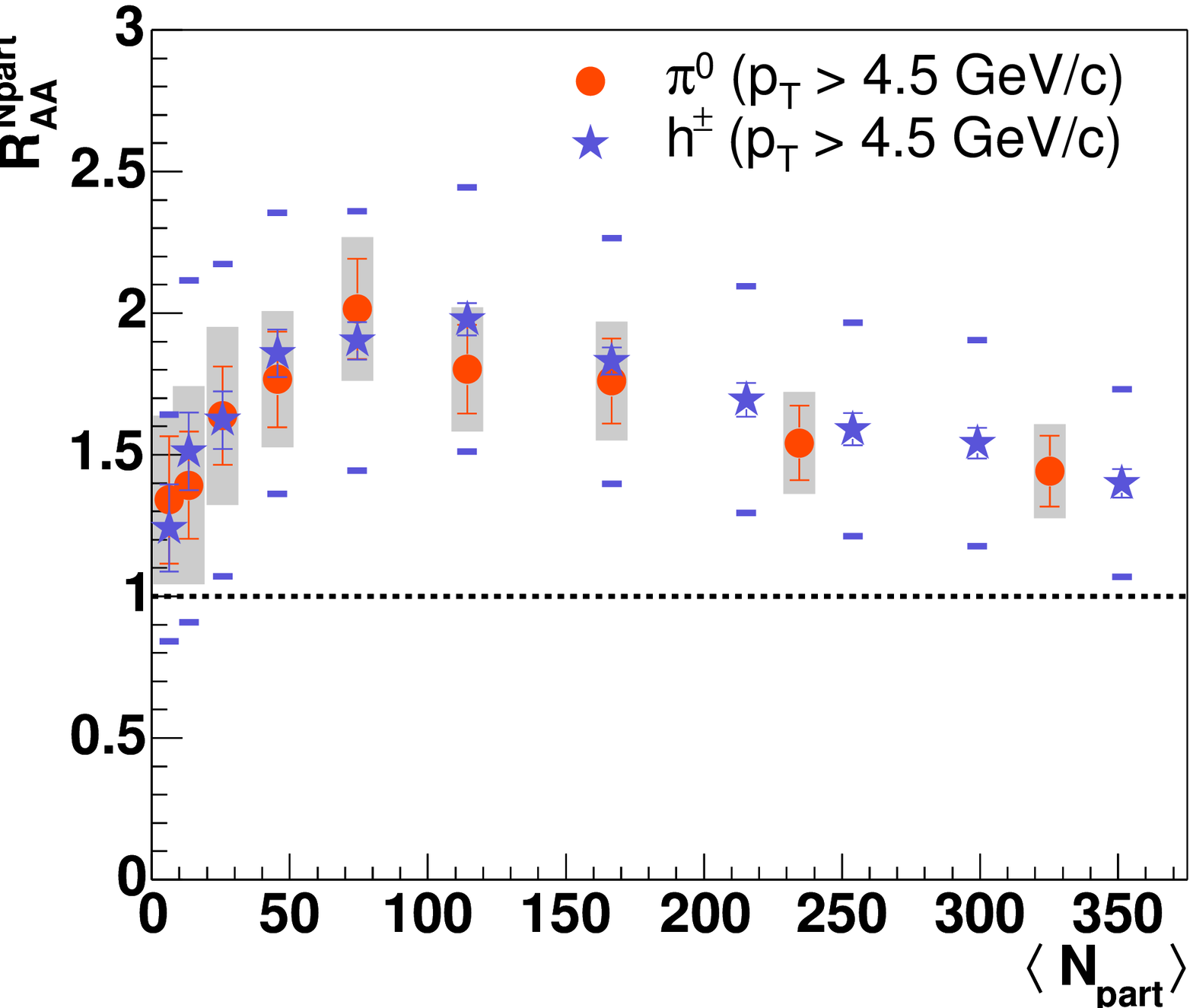}
\end{minipage}
\end{center}
\vspace*{-0.5cm}
\caption[]{{\it Left:} Ratio of central to semi-central $h^\pm$ yields normalized by $N_{part}$ 
in 4 different $p_{T}$ bins, as a function of $N_{part}$ measured by PHOBOS~\cite{phobos_hipt_200}.
The dashed (solid) line shows the expectation for $N_{coll}$ ($N_{part}$) scaling.
{\it Right:} Ratio of Au+Au over p+p $\pi^0$ and $h^\pm$ yields above 4.5 GeV/$c$
normalized by $N_{part}$ $(R_{AA}^{N_{part}})$ as a function of centrality given 
by $\langle N_{part} \rangle$ as measured by PHENIX ~\cite{phenix_pi0_200,phenix_hipt_200}.
The dashed line indicates the expectation for $N_{part}$ scaling.}
\label{fig:R_AA_Npart_vs_cent}
\end{figure}

$N_{part}$ (instead of $N_{coll}$) scaling at high $p_{T}$ is %(mind that $N_{part}^{pp}$ = 2) 
expected in scenarios dominated either by gluon saturation~\cite{dima} or by surface emission 
of the quenched jets~\cite{mueller}. ``Approximate'' $N_{part}$ scaling has been claimed by 
PHOBOS~\cite{phobos_hipt_200}: the ratio of central to a {\it fit} to {\it mid-central} 
yields in the range $p_{T}\approx$ 2. -- 4. GeV/$c$ stays flat as a function of 
centrality (Fig.~\ref{fig:R_AA_Npart_vs_cent}, left). However, at higher $p_{T}$ values, where
the suppression is seen to saturate at its maximum value, the centrality dependence of the ratio of 
$N_{part}$-scaled Au+Au over p+p yields for $\pi^0$ and $h^\pm$ measured by PHENIX 
(Fig.~\ref{fig:R_AA_Npart_vs_cent}, right) does not show a true participant scaling 
($R_{AA}^{part}>$ 1 for all centralities). Nonetheless, the fact that the production per participant 
pair above 4.5~GeV/{\it c} is, within errors, approximately constant over a wide range of 
intermediate centralities, is in qualitative agreement with a gluon saturation model 
prediction~\cite{dima}.

%%%%%%%%%%%%%%%%%%%%%%%%%%%%%%%%%%%%%%%%%%%%%%
%\clearpage

\subsection{High $p_{T}$ suppression: particle species dependence} %hadron composition}
\label{hadron_composition}

\begin{figure}[htbp]
\hspace*{-.8cm}
\begin{center}
\includegraphics[height=6.8cm]{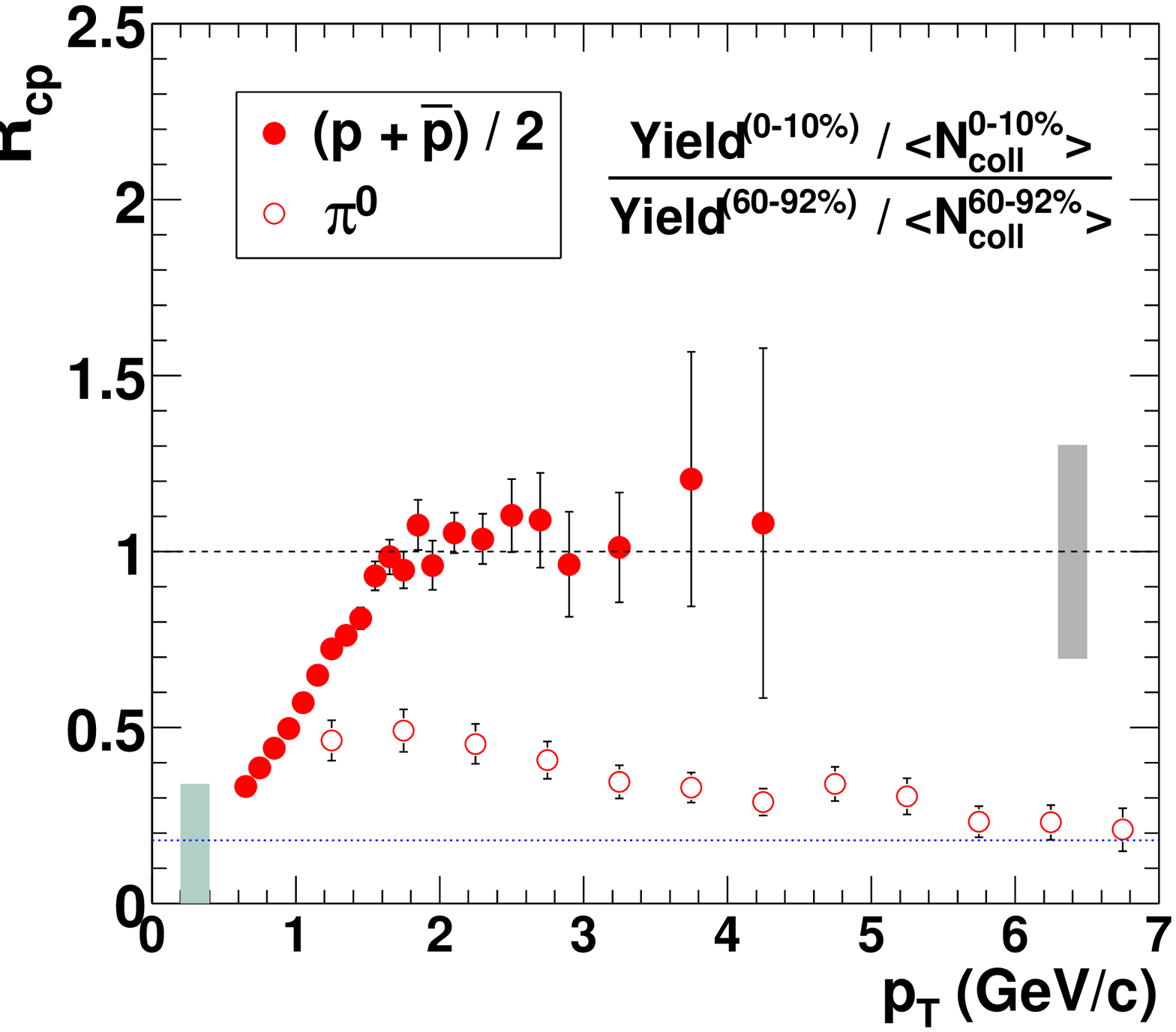}
\hspace*{4mm}
\includegraphics[height=7.cm]{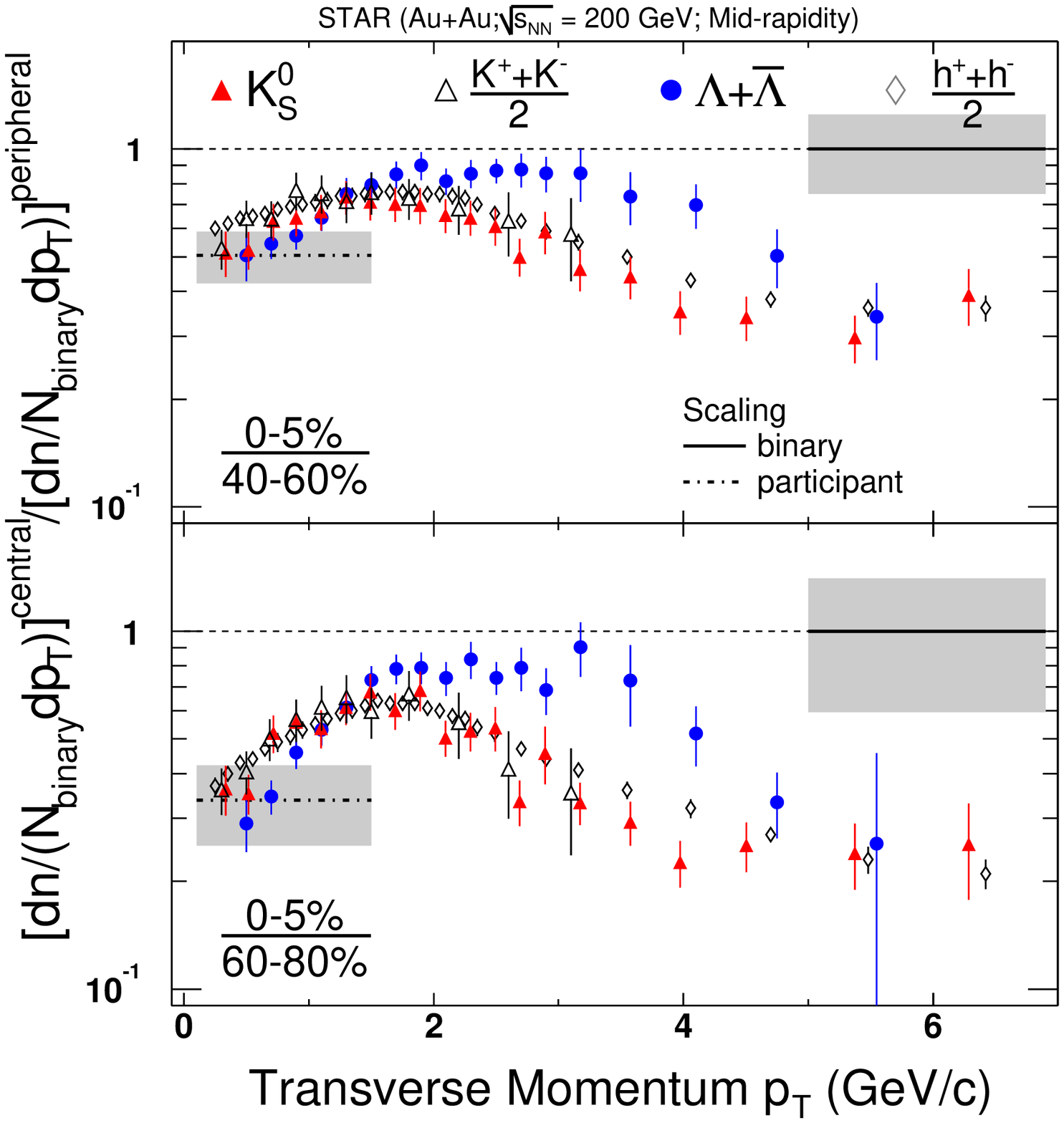}
\end{center}
%\hspace*{-2cm}
\vspace*{-0.7cm}
\caption[]{Ratio of central over peripheral $N_{coll}$ scaled yields, $R_{cp}(p_{T})$, 
as a function of $p_{T}$ for different species measured in Au+Au collisions: 
$(p+\bar{p})/2$ (dots) and $\pi^0$ (circles) by PHENIX~\protect\cite{phenix_ppbar_200} ({\it left}), 
and $\Lambda,\bar{\Lambda}$ (circles) and $K^0_s,K^\pm$ (triangles)
by STAR~\protect\cite{star_hipt_strange_200} ({\it right}).}
\label{fig:flavor_dep}
\end{figure}

One of the most intriguing results of the RHIC program so far is the different suppression 
pattern of baryons and mesons at moderately high $p_{T}$'s.
Figure~\ref{fig:flavor_dep} (left) compares the $N_{coll}$ scaled central to
peripheral yield ratios\footnote{Since the %(both the $\pi^0$ and the inclusive charged hadron~\cite{}) 
60--92\% peripheral Au+Au (inclusive and identified) spectra scale with $N_{coll}$ when 
compared to the p+p yields~\cite{phenix_pi0_200,star_hipt_200,phenix_ppbar_200}, 
$R_{cp}$ carries basically the same information as $R_{AA}$.} for $(p+\bar{p})/2$ and $\pi^0$: 
$R_{cp} = (yield^{(0-10\%)} / N_{coll}^{0-10\%}) / (yield^{(60-92\%)} /  N_{coll}^{60-92\%})$. 
From 1.5 to 4.5 GeV/$c$ the (anti)protons are not suppressed ($R_{cp}\sim$ 1) 
at variance with the pions which are reduced by a factor of 2 -- 3 in this $p_{T}$ range.
%Beyond $p_{T} \simeq$ 1.5 GeV/$c$ all spectra converge to the
%{\emph same} slope and seem to obey $N_{coll}$ scaling in agreement 
%with production due to hard processes in the absence of nuclear effects. 
If both $\pi^0$  and  $p,\bar{p}$ originate from the fragmentation 
of hard-scattered partons that lose energy in the medium, the nuclear 
modification factor $R_{ cp}$ should be independent of particle species 
contrary to the experimental result. The same discussion applies for
strange mesons and baryons as can be seen from the right plot of 
Fig.~\ref{fig:flavor_dep}. Whereas the kaon yields in central collisions 
are suppressed with respect to ``$N_{coll}$ scaling'' for 
all measured $p_{T}$, the yield of $\Lambda+\bar{\Lambda}$ is close to 
expectations from collision scaling in the $p_{T}$ range 1.8 -- 3.5 GeV/$c$. 
Interestingly, above $p_{T} \sim$ 5.0 GeV/$c$, the $K_S^0,K^\pm$, $\Lambda+\bar{\Lambda}$, 
and charged hadron yields are suppressed from binary scaling by a similar factor. 

\begin{figure}[htbp]
\vspace*{-.3cm}
\hspace*{-.8cm}
\begin{center}
\includegraphics[height=5.8cm]{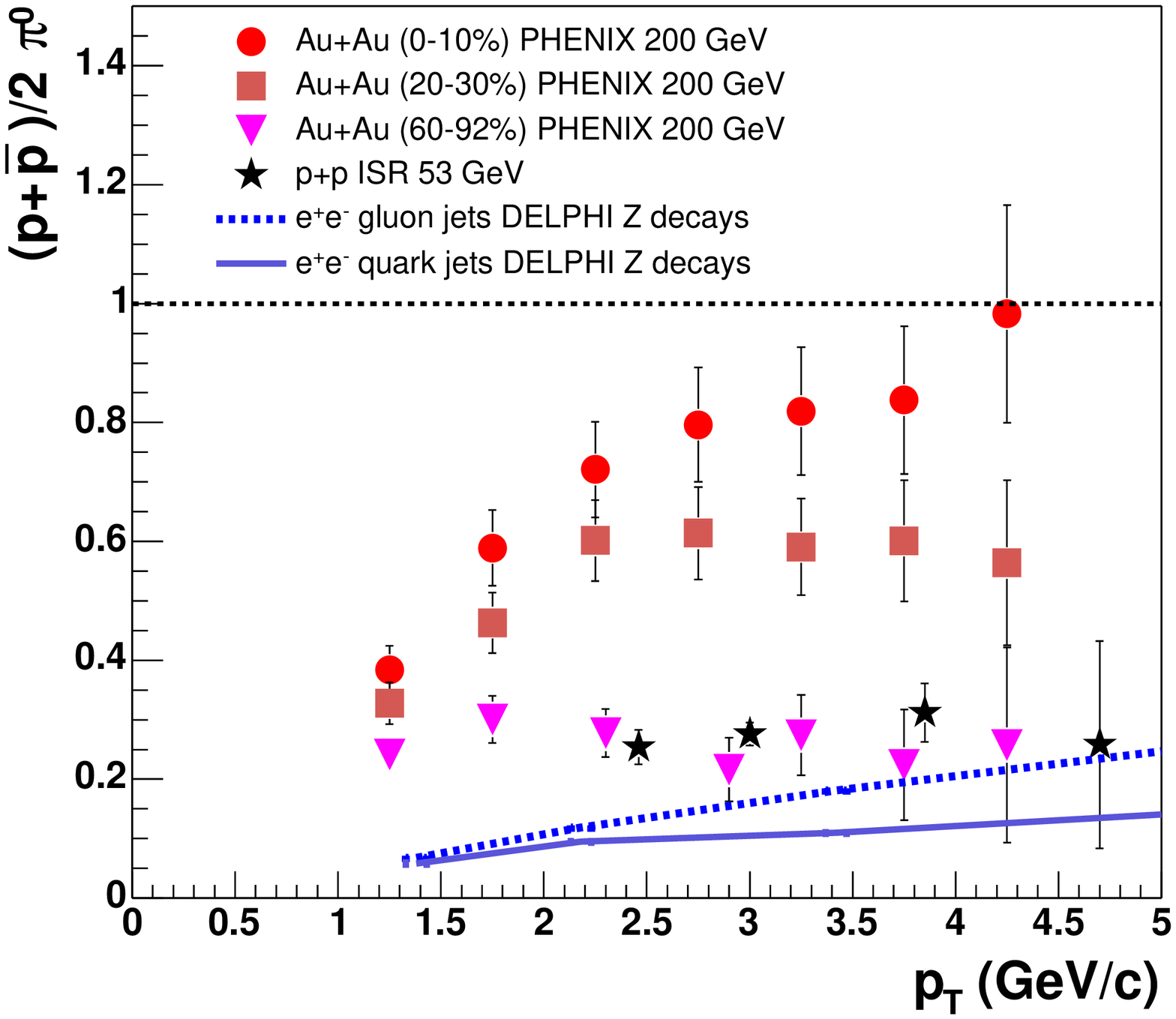}
\hspace*{4mm}
\includegraphics[height=6.0cm]{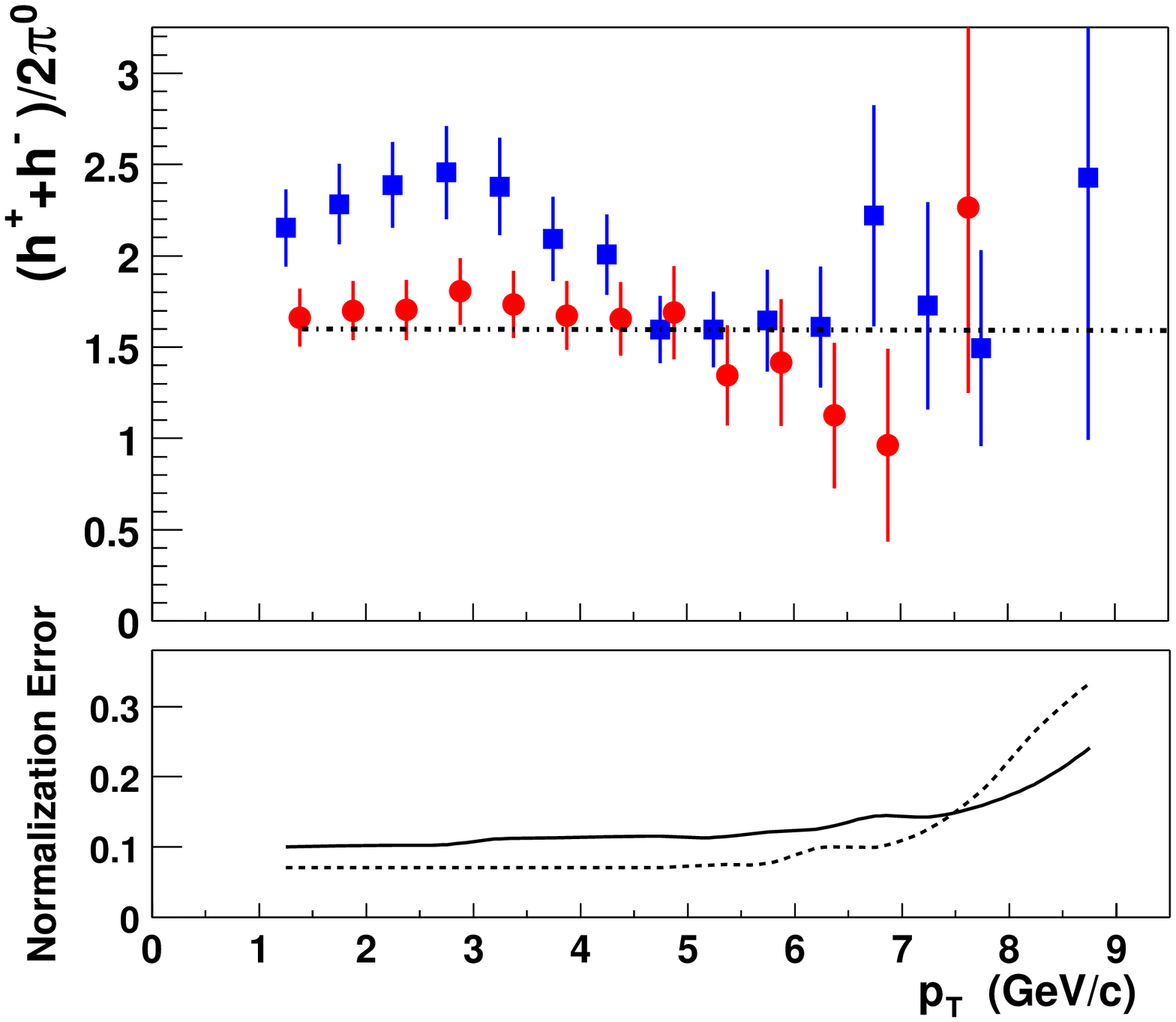}
\end{center}
%\hspace*{-2cm}
\vspace*{-0.7cm}
\caption[]{{\it Left}: Ratios of  $(p+\bar{p})/2$ over $\pi^0$ versus $p_{T}$ in central (dots), 
mid-central (squares), and peripheral (triangles) Au+Au, and in p+p (crosses), and 
$e^{+}e^{-}$ (dashed and solid lines) collisions~\cite{phenix_ppbar_200}.
{\it Right}: Ratio of charged hadron to $\pi^0$ in central (0--10\% -  squares) 
and peripheral (60--92\% - circles) Au+Au collisions compared to the 
$h/\pi\sim$ 1.6 ratio (dashed-dotted line) measured in p+p collisions~\cite{phenix_ppbar_200}.}
%The lower panel shows fractional normalization error common to both centrality selections (solid) 
%and the relative error between the two (dashed).}
\label{fig:flavor_dep2}
\end{figure}

Figure~\ref{fig:flavor_dep2} (left) shows the ratios of $(p+\bar{p})/2$ over $\pi^0$ as a
function of $p_{T}$ measured by PHENIX in central (0--10\%, circles), mid-central (20--30\%, squares), 
and peripheral (60--92\%, triangles) Au$+$Au collisions~\cite{phenix_ppbar_200}, 
together with the corresponding ratios measured in p+p collisions at 
CERN-ISR energies~\cite{ISR_ppbar,ISR_pi0} (crosses) and in gluon and quark jet fragmentation 
from $e^{+}e^{-}$ collisions~\cite{DELPHI} (dashed and solid lines resp.). Within errors, 
peripheral Au+Au results are compatible with the p+p and $e^{+}e^{-}$ ratios, but central 
Au+Au collisions have a $p/\pi$ ratio $\sim$ 4 -- 5 times larger. Such a result is at odds with 
standard perturbative production mechanisms, since in this case the particle ratios 
$\bar{p}/\pi$ and $p/\pi$ should be described by a universal fragmentation function independent 
of the colliding system, which favors the production of the lightest particle.
Beyond $p_{T} \approx 4.5$~GeV/$c$, the identification of charged particles is not yet
possible with the current PHENIX configuration, however the measured $h/\pi^0\sim$~1.6
ratio above $p_{T}\sim$ 5 GeV/$c$ in central and peripheral Au+Au 
is consistent with that measured in p+p collisions (Fig. \ref{fig:flavor_dep2}, right).
This result together with STAR $R_{cp}$ result on strange hadrons (Fig.~\ref{fig:flavor_dep}, right)
supports the fact that for large $p_{T}$ values the properties of the baryon production mechanisms 
approach the (suppressed) meson scaling, thus limiting the observed baryon enhancement in central 
Au$+$Au collisions to the intermediate transverse momenta $p_{T} \lesssim$ 5 GeV/$c$.

Several theoretical explanations (see refs. in ~\cite{phenix_ppbar_130,phenix_ppbar_200}) 
have been proposed to justify the different behaviour of mesons and baryons at intermediate 
$p_{T}$'s based on: (i) quark recombination (or coalescence), (ii) medium-induced difference in the 
formation time of baryons and mesons, (iii) different ``Cronin enhancement'' for protons
and pions, or (iv) ``baryon junctions''.
In the recombination picture %(see Section 4.41 of this Report and 
~\cite{recomb} the partons from a thermalized system coalesce and 
with the addition of quark momenta, the soft production of baryons extends to much 
larger values of $p_{T}$ than that for mesons.  In this scenario, the effect is
limited to $p_{T} < 5$\,GeV, beyond which fragmentation becomes the
dominant production mechanism for all species.

%%%%%%%%%%%%%%%%%%%%%%%%%%%%%%%%%%%%%%%%%%%%%%
%\clearpage

\subsection{High $p_{T}$ suppression: pseudorapidity dependence}

\begin{figure}[htbp]
\begin{center}
\begin{minipage}[t]{75mm}
\includegraphics[height=5.2cm]{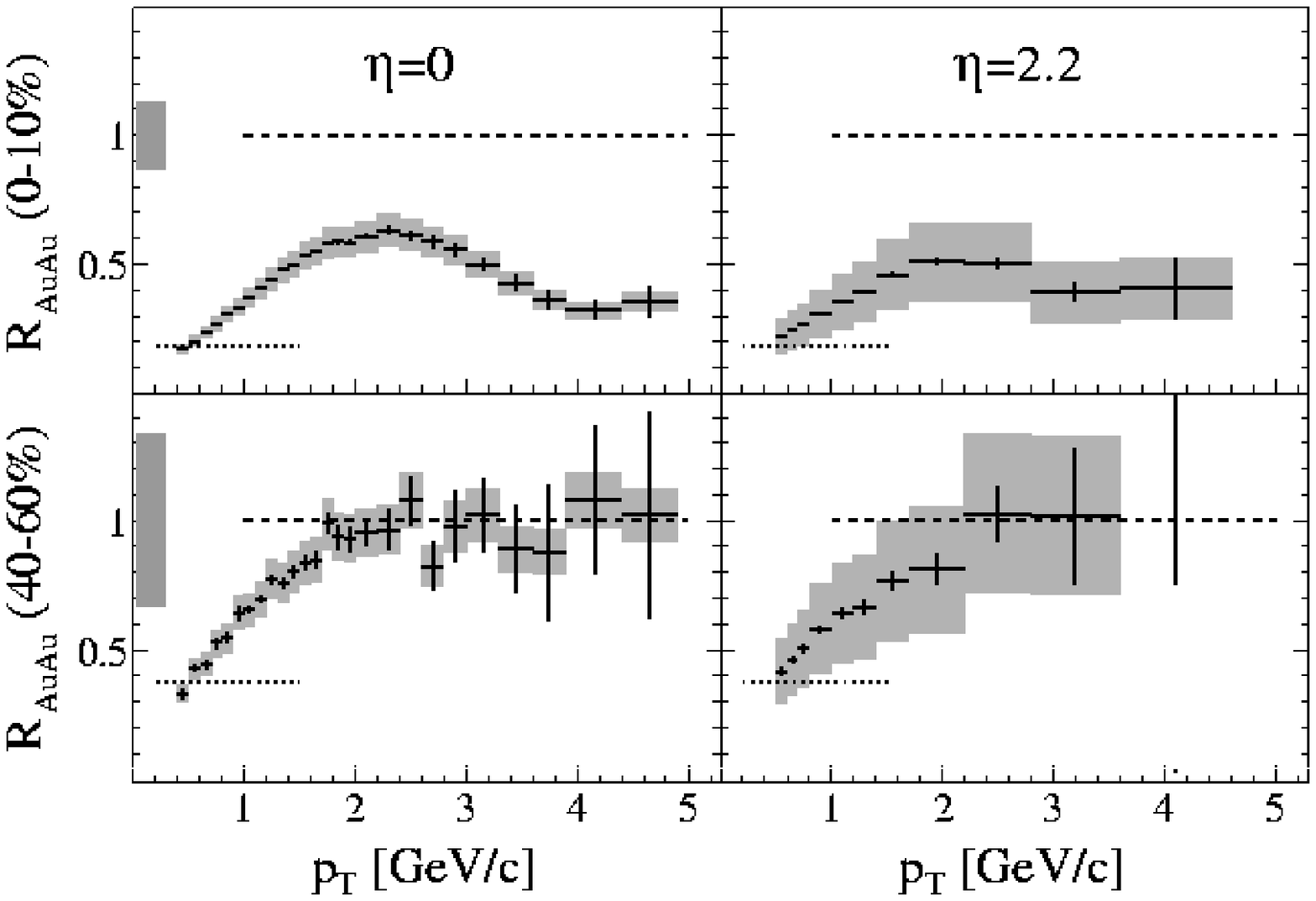}
\end{minipage}
\hspace*{0.5cm}
\begin{minipage}[t]{75mm}
\includegraphics[height=5.4cm]{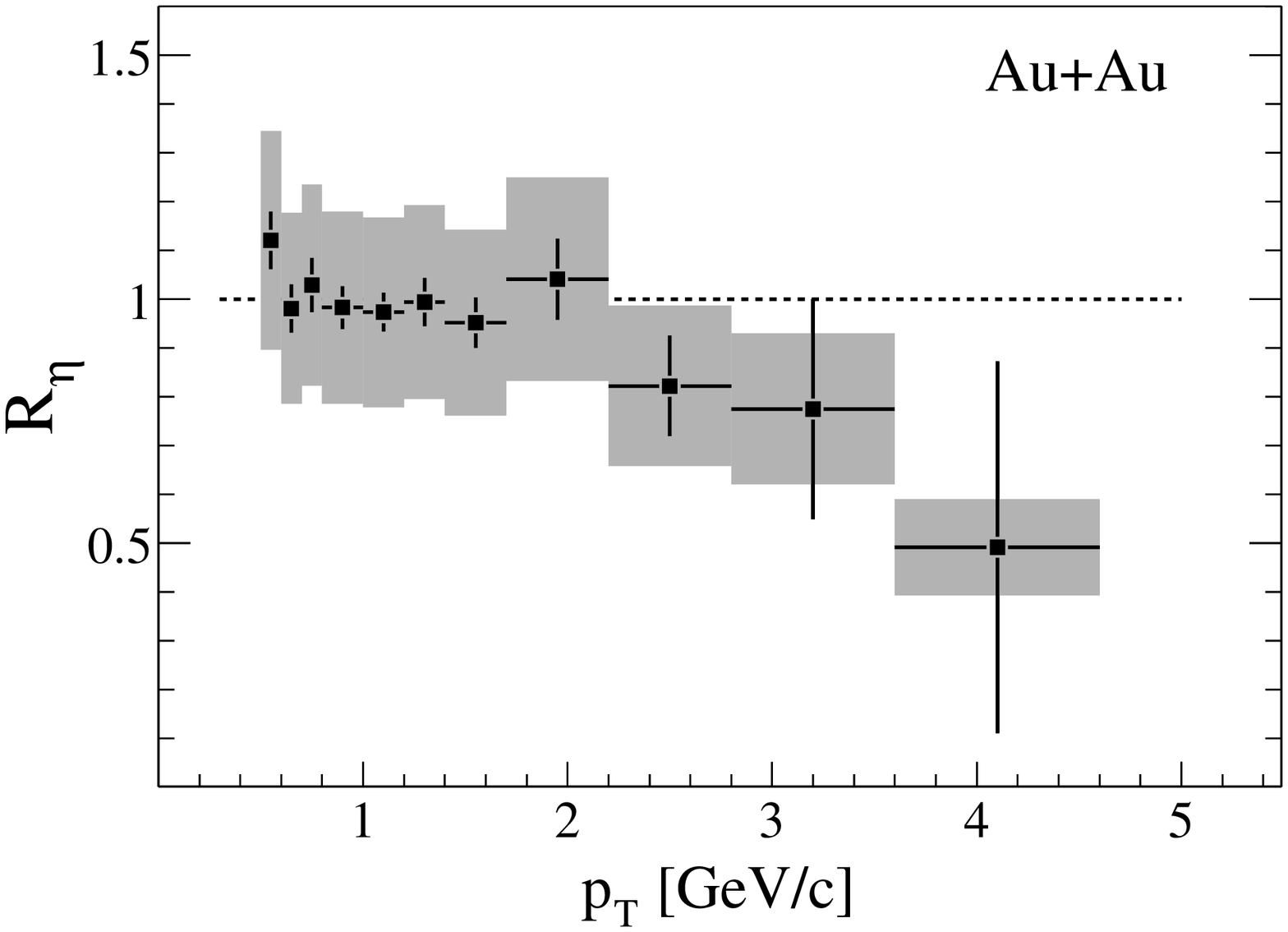}
\end{minipage}
\end{center}
\vspace*{-0.3cm}
\caption{{\it Left}: $R_{AA}(p_{T})$  measured by BRAHMS at $\eta$ = 0 and $\eta$ = 2.2 for 0--10\% 
most central and for semi-peripheral (40-60\%) Au+Au collisions.
{\it Right}: Ratio $R_\eta$ of $R_{cp}$ distributions at $\eta=2.2$ and $\eta=0$. 
Figs. from ~\protect\cite{brahms_hipt_200}.}
\label{fig:brahms_RAA}
\end{figure}

BRAHMS is, so far, the only experiment at RHIC that has measured high $p_{T}$ inclusive 
charged hadron spectra off mid-rapidity. Fig. \ref{fig:brahms_RAA} (left) shows the nuclear
modification factors $R_{AA}(p_{T})$ for central and semi-peripheral Au+Au measurements 
at mid-pseudorapidity ($\eta$ = 0) and at $\eta$ = 2.2~\cite{brahms_hipt_200}.
The high $p_{T}$ suppression is not limited to central rapidities
but it is clearly apparent at forward $\eta$'s too. 
%Within errors no difference is apparent between both measurements.
Fig. \ref{fig:brahms_RAA} (right) %provides further insights into the pseudorapidity dependence of the suppression via 
shows the ratio of suppressions at the two pseudorapidity values, $R_\eta = R_{cp}(\eta=2.2)/R_{cp}(\eta=0)$.
The high $p_{T}$ deficit at $\eta$ = 2.2 is similar to, or even larger than, at $\eta$ = 0 
indicating that the volume causing the suppression extends also in the longitudinal direction. 
%These results further constraint the mechanisms of high $p_{T}$ production in Au+Au.

%%%%%%%%%%%%%%%%%%%%%%%%%%%%%%%%%%%%%%%%%%%%%%
%\clearpage

\section{High $p_{T}$ azimuthal correlations in Au+Au collisions}

There are two main sources of azimuthal correlations at high $p_{T}$ in heavy-ion 
collisions:

\begin{itemize}
\item The {\it fragmentation} of hard-scattered {\it partons} results in jets of 
high $p_{T}$ hadrons correlated in both rapidity and azimuthal angle. Such correlations
are short range ($\Delta\eta\lesssim$ 0.7, $\Delta\phi\lesssim$ 0.75), involve comparatively 
large transverse momentum particles ($p_{T}>$ 2 GeV/$c$), and are unrelated 
(in principle) to the orientation of the $AA$ reaction plane.

\item {\it Collective (elliptic) flow}: % resulting from t
The combination of (i) the geometrical asymmetry in non-central $AA$ reactions 
(``almond''-like region  of the overlapping nuclei), and (ii) multiple reinteractions 
between the produced particles in the overlap region; results in %Those two effects generate
pressure gradients in the collision ellipsoid which transform the original
coordinate-space asymmetry into a momentum-space anisotropy. 
The amount of elliptic flow (a true collective effect absent
in p+p collisions) is measured by the second harmonic coefficient, 
$v_{2}\equiv \langle cos(2\phi)\rangle$, of the Fourier expansion 
of the particles azimuthal distribution with respect to the reaction plane.
\end{itemize}

Additionally, there are other second-order sources of angular correlations
like resonance decays, final state (particularly Coulomb) interactions, 
momentum conservation, or other experimental effects like photon conversions,
which have to be subtracted out in order to extract the interesting ``jet-like''
or ``flow-like'' signals.

\subsection{High $p_{T}$ azimuthal correlations: Jet signals}

\begin{figure}[htbp]
%\center
\includegraphics[width=.45\textwidth]{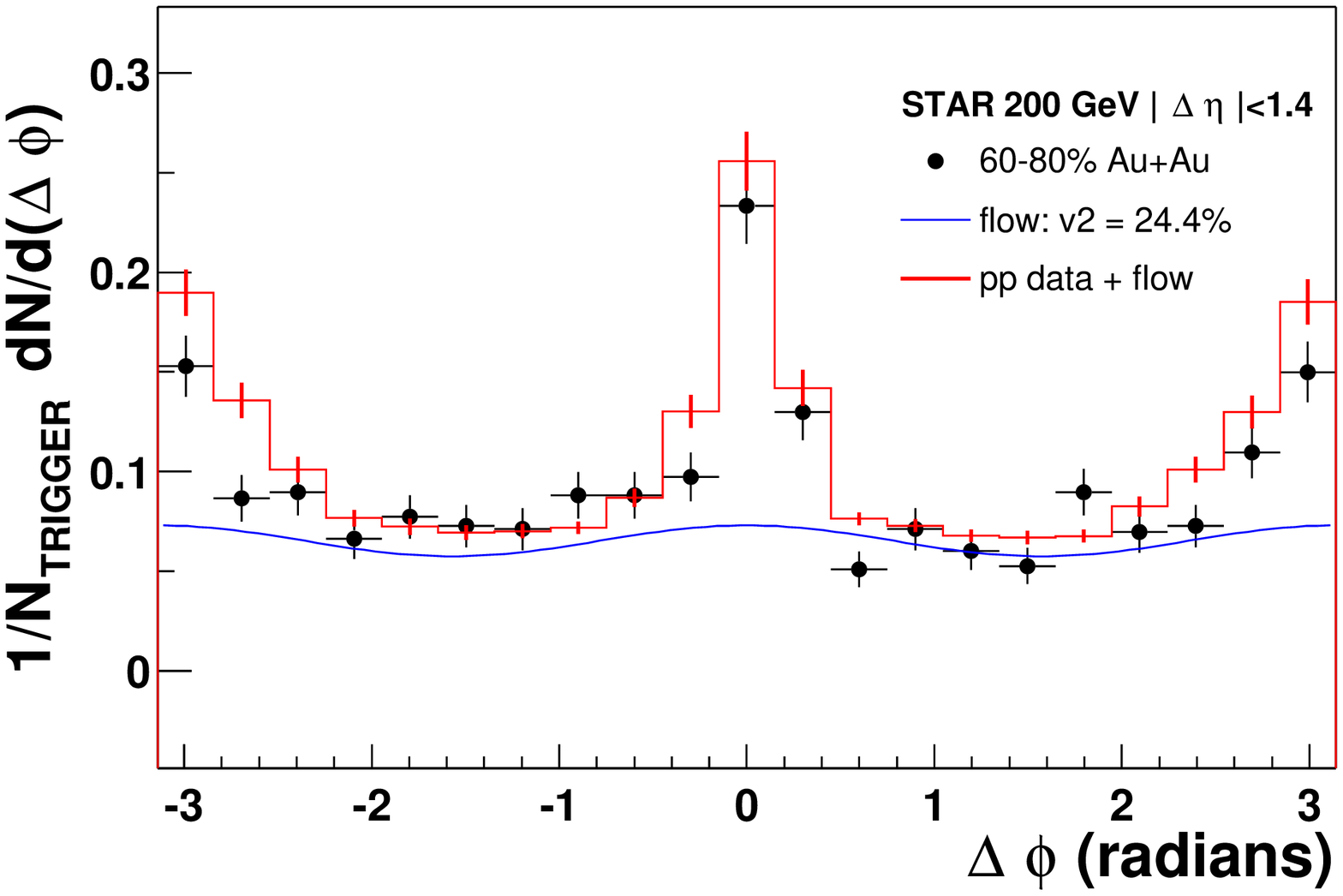}
\hspace{.1\textwidth}
\includegraphics[width=.45\textwidth]{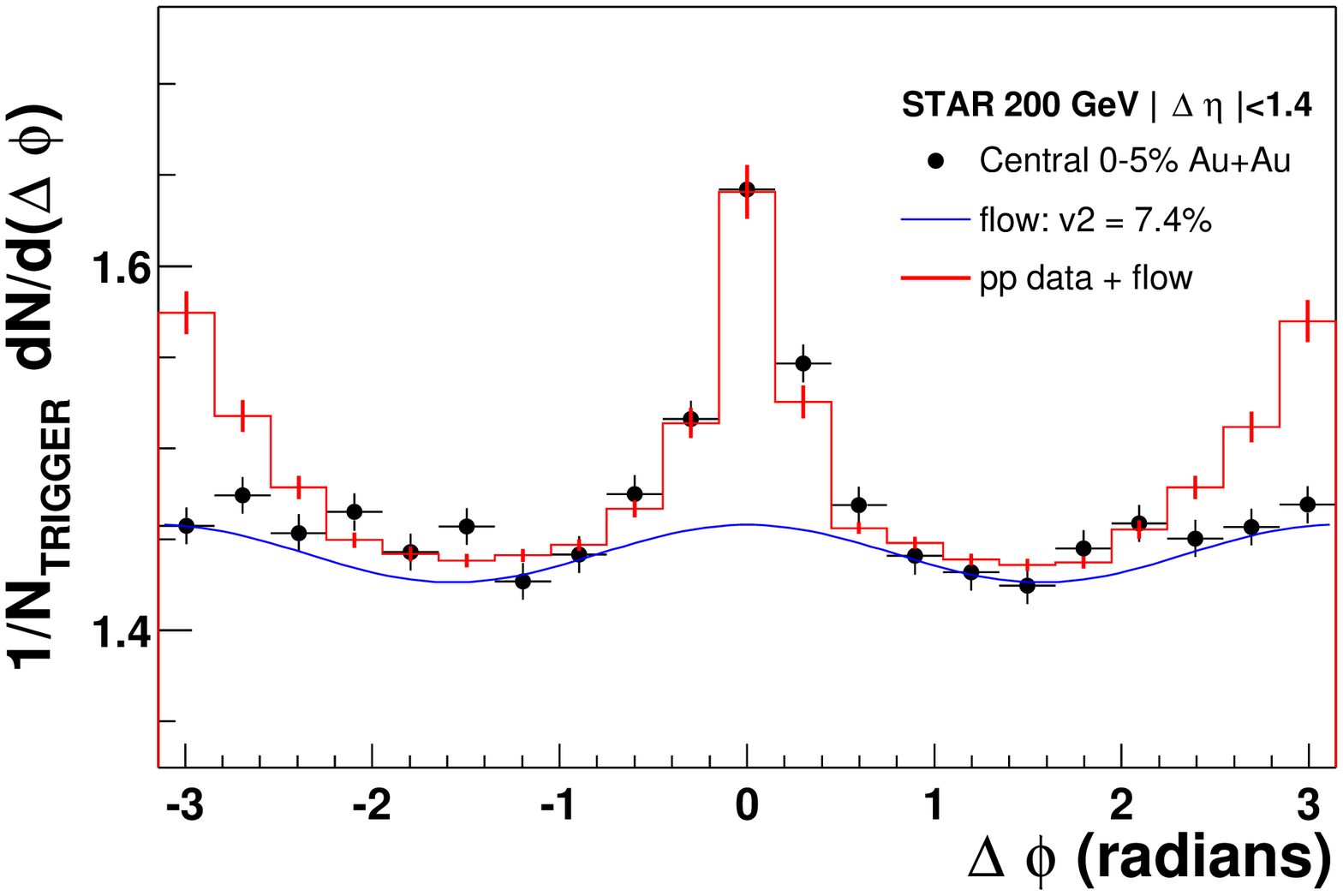}
\caption{Azimuthal correlations for peripheral ({\it left}) and central ({\it right}) 
Au+Au collisions compared to the pedestal and flow-scaled correlations
in p+p collisions. Fig. from ~\cite{jacobs_klay}.}
\label{fig:jets_azim_corr_star}
\end{figure}

Although, standard jet reconstruction algorithms fail below $p_{T}\approx$ 40 GeV/$c$ 
when applied to the soft-background dominated environment of heavy-ion 
collisions, angular correlations of pairs of high $p_{T}$ particles~\cite{star_away_side,chiu_qm02}
have been very successfully used to study on a statistical basis the properties 
of the produced jets. For each event with ``trigger'' particle(s) with
$p_{T}$ = 4 -- 6 GeV/$c$ and ``associated'' particle(s) with $p_{T}$ = 2 -- 4 GeV/$c$
and $|\eta|<$ 0.7, STAR~\cite{star_away_side} determines the two-particle azimuthal 
distribution
\begin{equation}
D(\Delta \phi) \propto \frac{1}{N_{trigger}}%\frac{1}{\epsilon}
\frac{dN}{d(\Delta\phi)}\;.
\end{equation}
%(where $\epsilon$ is the tracking efficiency of the associated particles).
Fig.~\ref{fig:jets_azim_corr_star} shows $D(\Delta \phi)$ for peripheral (left) and
central (right) Au+Au collisions (dots) compared to $D(\Delta \phi)$ from p+p collisions
(histogram), and to a superposed $cos(2\Delta\phi)$ flow-like term (blue curve). 
On the one hand, the correlation strength at small relative angles ($\Delta \phi\sim$ 0) 
in peripheral and central Au+Au as well as at back-to-back angles ($\Delta \phi\sim \pi$) 
in peripheral Au+Au are very similar to the scaled correlations in p+p collisions. 
The near-side peaks in all three collision systems are characteristic of jet 
fragmentation~\cite{star_away_side} (a result also observed by PHENIX using {\it neutral} 
trigger particles~\cite{chiu_qm02}). On the other hand, the away-side peak 
($\Delta \phi\sim \pi$) in central collisions is strongly suppressed.

\begin{figure}[htbp]
\begin{center}
\includegraphics[height=6.cm]{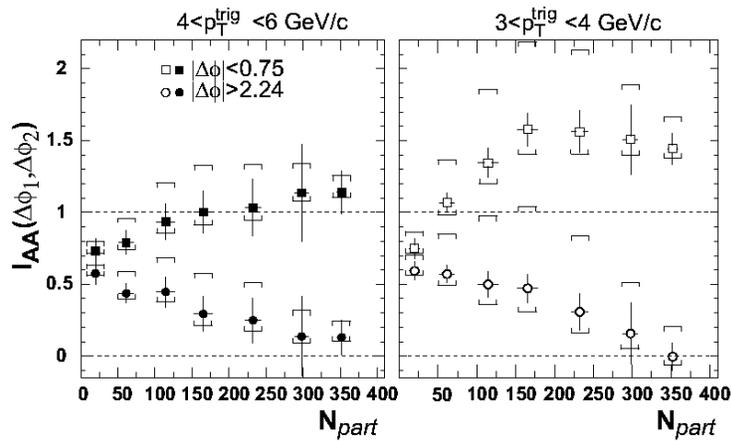}
\end{center}
\vspace*{-0.1cm}
\caption[]{Ratio of Au+Au over p+p integrated azimuthal correlations, 
Eq. (\protect{\ref{eq:I_AA}}), for small-angle (squares, $|\Delta \phi|<0.75$ 
radians) and back-to-back (circles, $|\Delta \phi|>2.24$ radians) azimuthal regions 
versus number of participating nucleons 
for trigger particle intervals $4<p_{T}^{trig}<6$ GeV/$c$ (solid) and
$3<p_{T}^{trig}<4$ GeV/$c$ (hollow)~\cite{star_away_side}.}
\label{fig:I_AA_star}
\end{figure}

In order to study the evolution as a function of centrality of the
the near-side, $D^{AuAu}(\Delta \phi<0.75)$, and away-side, $D^{AuAu}(\Delta \phi>2.24)$,
angular correlations in Au+Au compared to p+p, $D^{pp}$, 
STAR has constructed the quantity
\begin{eqnarray}
I_{AA}(\Delta \phi_1,\Delta \phi_2) = \frac{\int_{\Delta \phi_1}^{\Delta \phi_2} d(\Delta \phi) [D^{\mathrm{AuAu}}- B(1+2v_{2}^2 \cos(2 \Delta \phi))]}{\int_{\Delta \phi_1}^{\Delta \phi_2} d(\Delta \phi) D^{\mathrm{pp}}},
\label{eq:I_AA}
\end{eqnarray}
where $B$ accounts for overall background and $v_{2}$ the azimuthal correlations due to elliptic flow.
Fig.~\ref{fig:I_AA_star} shows $I_{AA}$ for the near-side (squares) and
away-side (circles) correlations as a function of
the number of participating nucleons ($N_{part}$). On the one hand,
the near-side correlation function is relatively suppressed compared to
the expectation from Eq. (\ref{eq:I_AA}) in the most peripheral region 
(a result not completely understood so far) and increases slowly with $N_{part}$.
On the other hand, the back-to-back correlation strength above the background 
from elliptic flow, decreases with increasing $N_{part}$ 
and is consistent with zero for the most central collisions.
The disappearance of back-to-back jet-like correlations is consistent 
with large energy loss effects in a system that is opaque to the propagation 
of high-momentum partons or their fragmentation products.   

%%%%%%%%%%%%%%%%%%%%%%%%%%%%%%%%%%%%%%%%%%%%%%
%\clearpage

\subsection{High $p_{T}$ azimuthal correlations: Collective elliptic flow}

\begin{figure}[htbp]
\begin{center}
%\hspace*{-1.cm}
\includegraphics[height=4.cm]{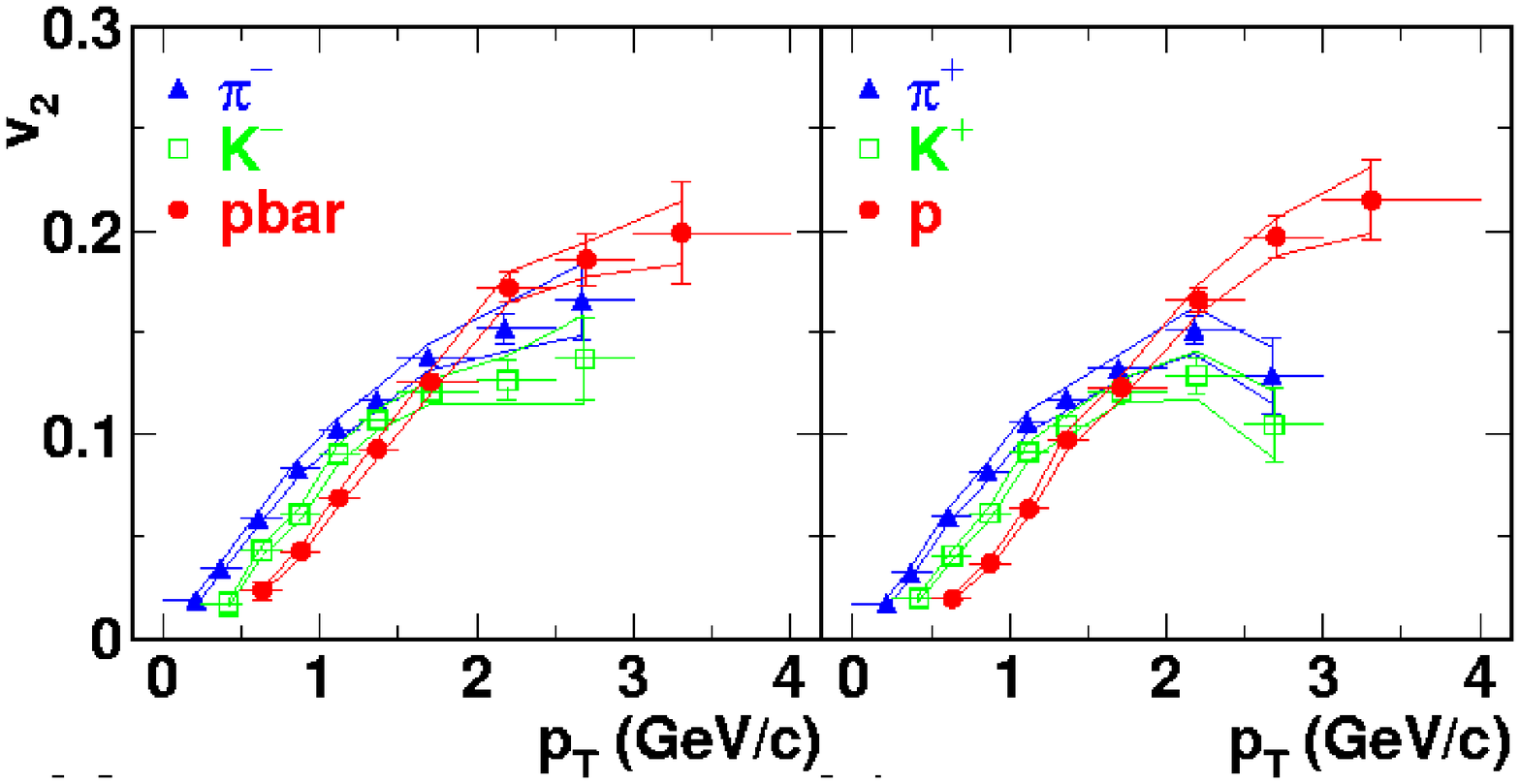}
\hspace*{0.5cm}
\includegraphics[height=4.2cm]{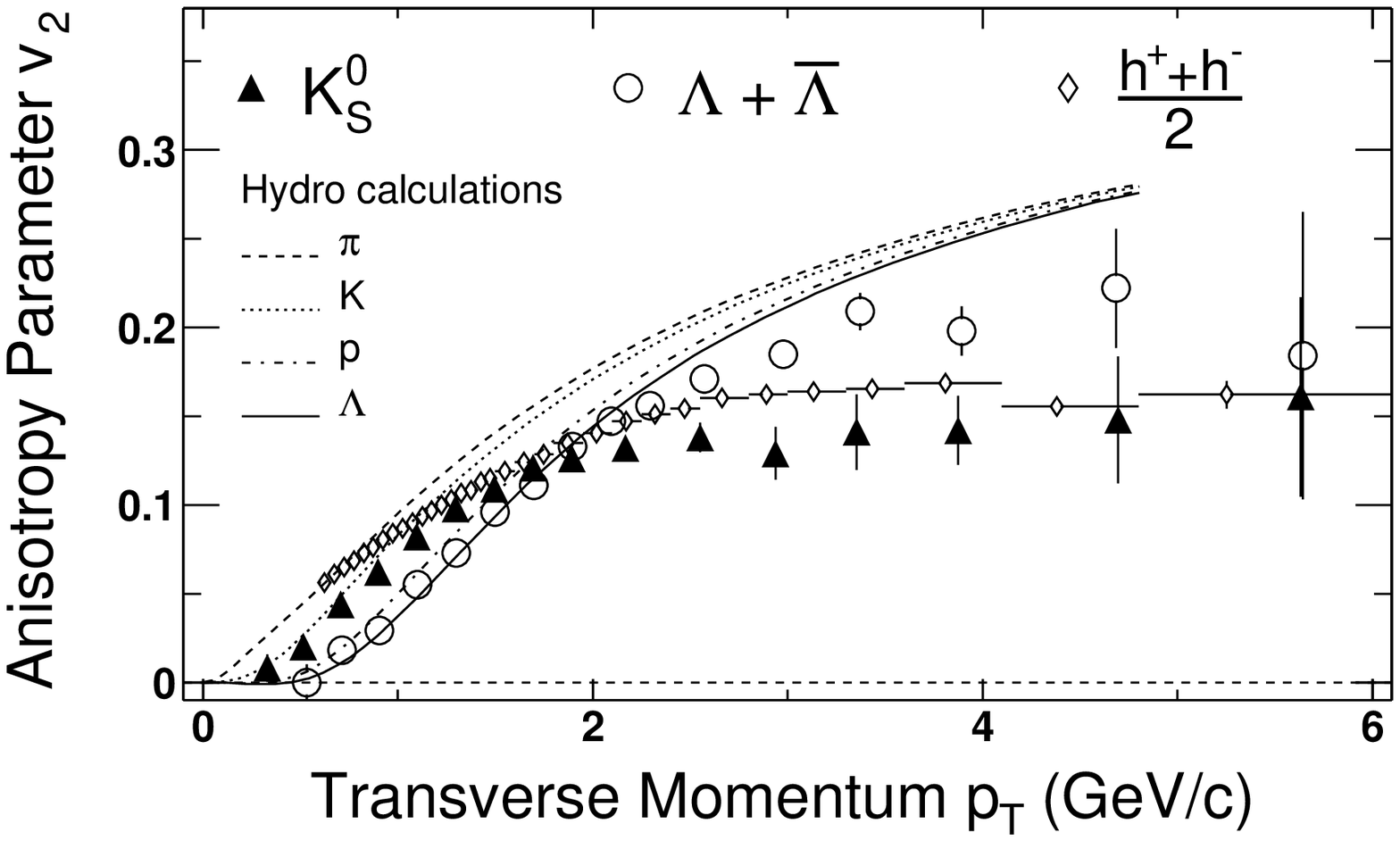}
\vspace*{-0.5cm}
\end{center}
\caption[]{$v_{2}$ as a function of transverse momentum for identified particles at RHIC:
$\pi^\pm$, $K^\pm$ and $p,\bar{p}$ from PHENIX ({\it left}), and $K^0_s$ and $\Lambda,\bar{\Lambda}$
from STAR ({\it right}).}
\label{fig:v2_pT}
\end{figure}

At low $p_{T}$ the strength of the elliptic flow signal is found to be large and 
consistent with hydrodynamics expectations. Above $p_{T}\sim$ 2 GeV/$c$
where the contribution from collective behaviour is negligible, $v_{2}$ 
is found to be still a sizeable signal which saturates and/or slightly decreases
as a function of $p_{T}$~\cite{star_hipt_flow_130,phenix_flow,star_hipt_strange_200}. 
The large value $v_{2}(p_{T}>$ 2 GeV/$c) \sim$ 0.15 implies unrealistically large 
parton densities and/or cross-sections according to standard parton 
transport calculations~\cite{recomb_flow}. Various interpretations have been proposed 
to account for such a large $v_{2}$ parameter within different physical scenarios. 
In jet quenching models~\cite{glv_flow} the resulting momentum anisotropy 
results from the almond-like density profile of the opaque medium (see, however,~\cite{shuryak_flow}). 
Calculations based on gluon saturation~\cite{cgc_flow} yield a (``non-flow'') azimuthal 
asymmetry component from the fragmentation of the released gluons from the initial-state 
saturated wave functions of the colliding nuclei. Finally, quark recombination 
effects~\cite{recomb_flow} can naturally enhance the elliptic flow of the produced hadrons 
compared to that of partons. The measured $v_{2}(p_{T})$ for mesons and baryons 
shows a distinct pattern (Fig. \ref{fig:v2_pT}): $v_{2}^{m}>v_{2}^{b}$ at 
low $p_{T}$, $v_{2}^{m}\approx v_{2}^{b}$ at $p_{T}\approx$ 2 GeV/$c$, and
$v_{2}^{m}<v_{2}^{b}$ at higher $p_{T}$'s; which further constraints the
proposed theoretical explanations.

\begin{figure}[htbp]
\begin{center}
%\hspace*{-0.2cm}
\includegraphics[height=4.cm]{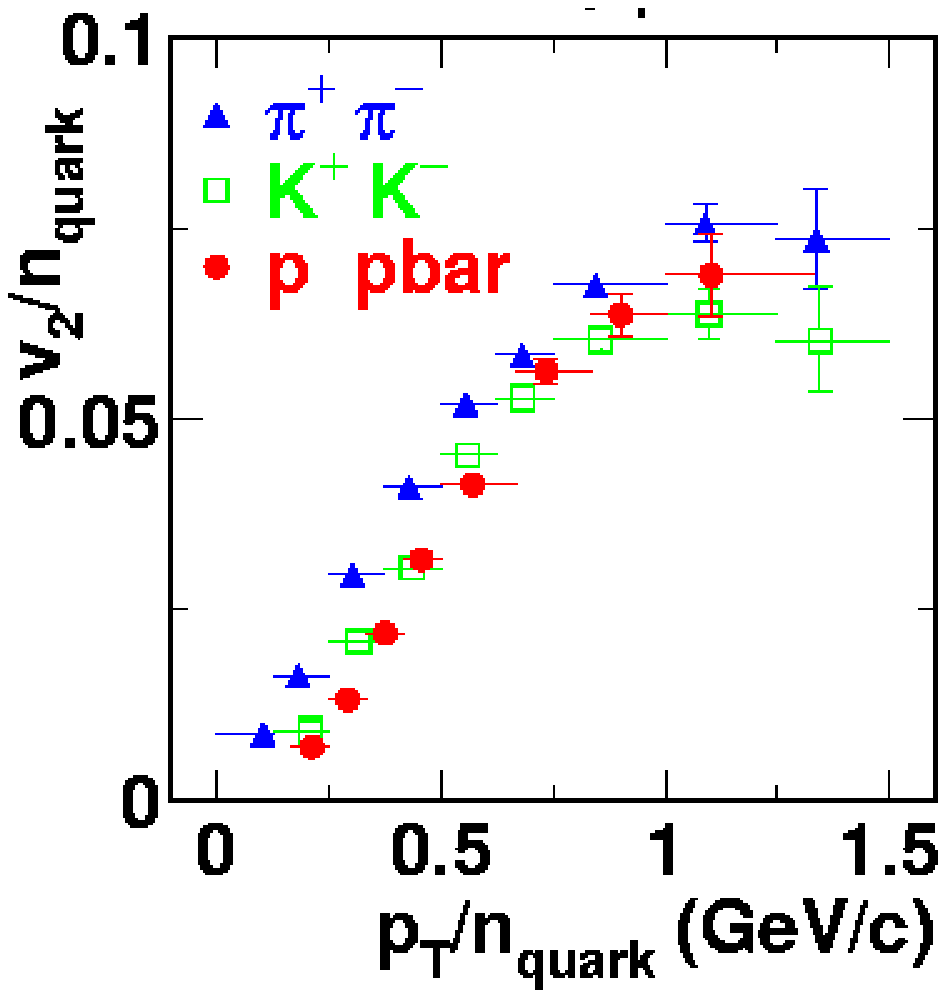}
\hspace*{1.0cm}
\includegraphics[height=4.2cm]{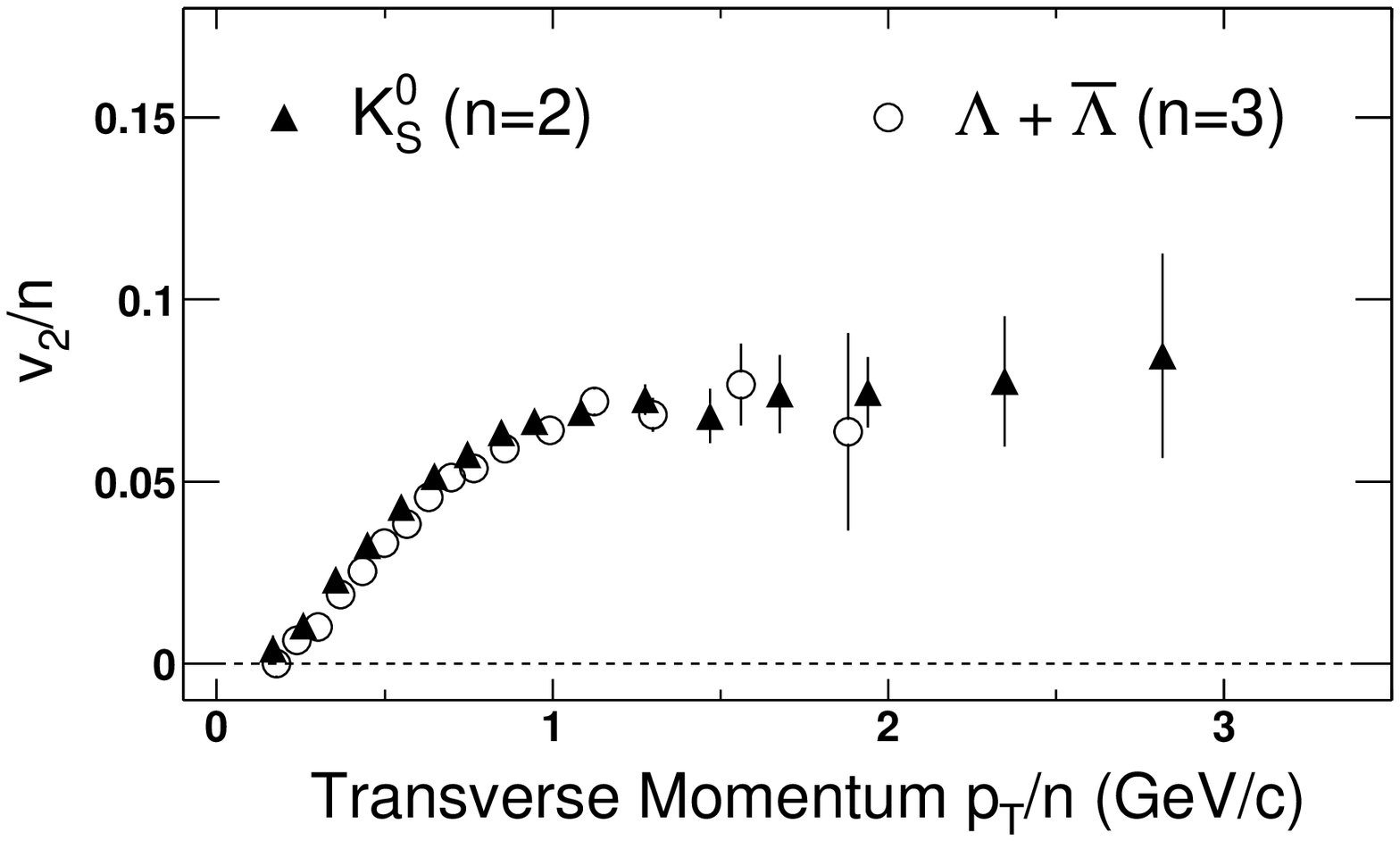}
\vspace*{-0.3cm}
\end{center}
\caption[]{The $v_{2}$ parameter scaled by the number of constituent
quarks ($n$) versus $p_{T}/n$ for $\pi^\pm$, $K^\pm$ and $p,\bar{p}$ 
(PHENIX \cite{phenix_flow}, {\it left}) and $K^0_s$ and $\Lambda,\bar{\Lambda}$ 
(STAR \cite{star_hipt_strange_200}, {\it right}).}
\label{fig:v2_n_pT}
\end{figure}

Quark coalescence models~\cite{recomb_flow} naturally lead to weaker
baryon flow than meson flow at low $p_{T}$, while the opposite holds
above 2 GeV/$c$. This simple mass ordering expectation %prediction from recombination models 
is confirmed by the identified particle data from PHENIX
and STAR (Fig. \ref{fig:v2_n_pT}). The fact that the $v_{2}$ parameters 
scaled by the number of constituent quarks ($n$ = 2 for mesons, $n$ = 3 baryons) 
versus $p_{T}/n$, globally fall in a single curve, supports the scenario where
hadrons at moderate $p_{T}$'s form by coalescence of co-moving quarks.

%%%%%%%%%%%%%%%%%%%%%%%%%%%%%%%%%%%%%%%%%%%%%%
\clearpage

\section{High $p_{T}$ hadron production in d+Au collisions}

\begin{figure}[htbp]
\vspace*{-0.1cm}
\begin{center}
\begin{tabular}{cc}
   \includegraphics[height=4.2cm]{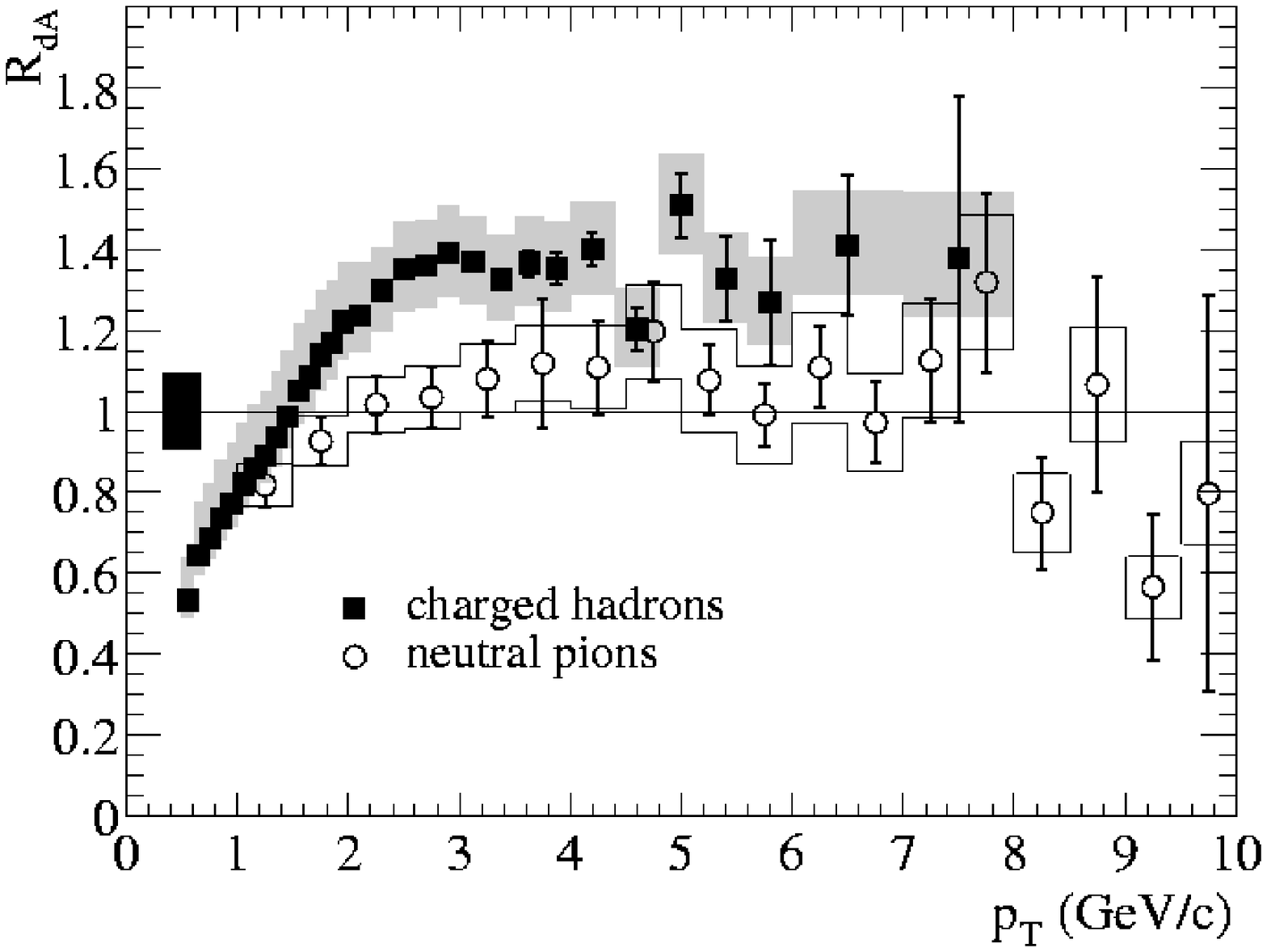} &
   \includegraphics[height=4.2cm]{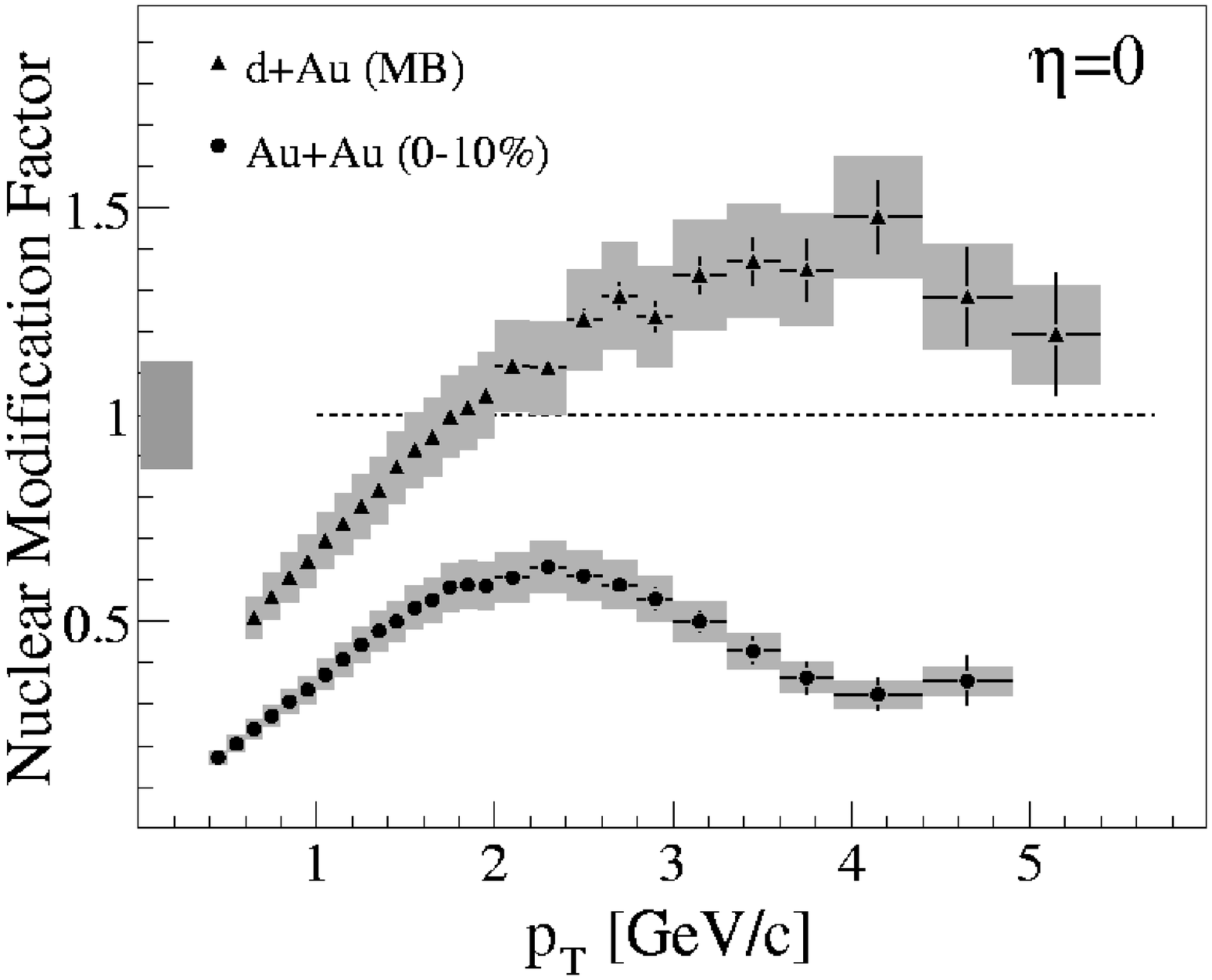} \\
   \includegraphics[height=5.0cm]{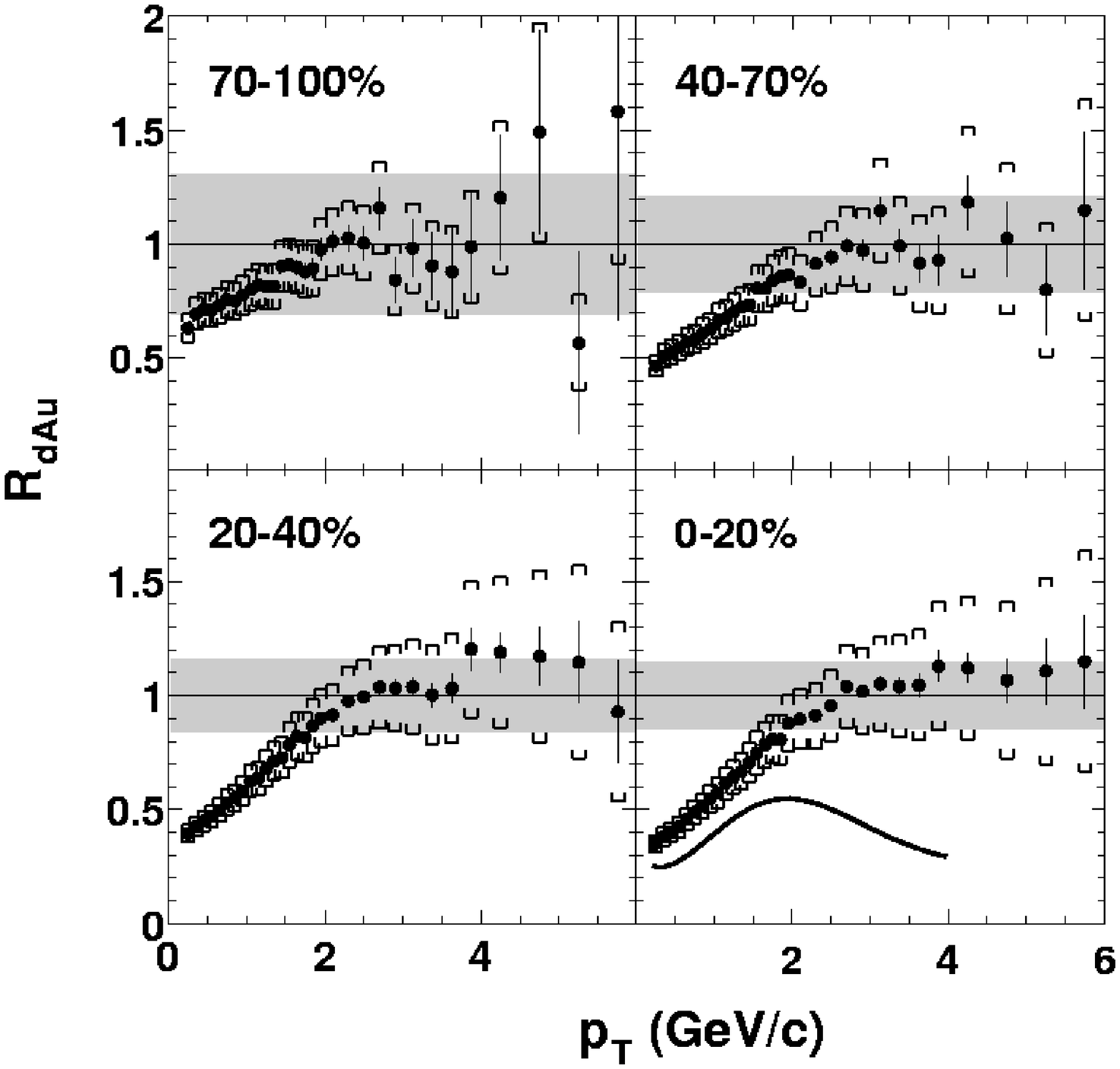} &
   \includegraphics[height=4.cm]{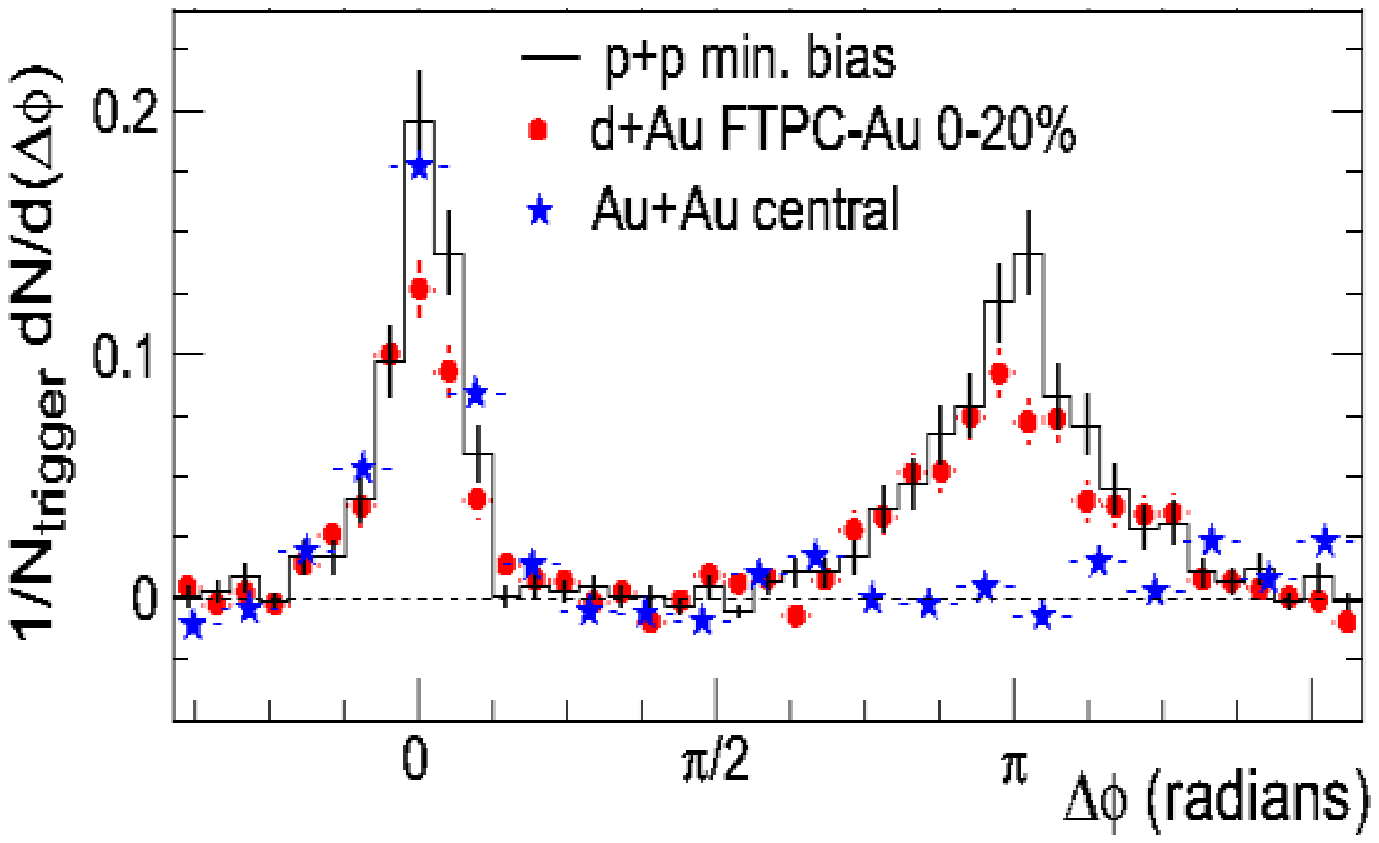} \\
\end{tabular}
\end{center}
\vspace*{-0.5cm}
\caption[]{Top: Nuclear modification factor $R_{dAu}$ for MB d+Au: $h^\pm$ and $\pi^0$ (PHENIX, 
{\it left}), $h^\pm$ (BRAHMS, {\it right}). Bottom: $R_{dAu}$ for $h^\pm$
measured by PHOBOS in 4 different d+Au centralities ({\it left}), and comparison of two-particle 
azimuthal distributions for central d+Au, p+p, and central Au+Au collisions (STAR, {\it right}).}
\label{fig:dAu}
\end{figure}

Proton- (or deuteron-) nucleus collisions constitute a reference ``control'' experiment  
needed to determine the influence of {\it cold} nuclear matter effects in high $p_{T}$ 
hadro-production. Since final-state medium effects are marginal in p,d+Au collisions, 
they are basic tools to ascertain whether models based on initial- or 
final- state QCD effects can explain the distinct hard scattering behaviour 
observed in Au+Au collisions at RHIC. During the third year of RHIC operation, the 4
experiments collected data from d+Au collisions at $\sqrt{s_{_{NN}}}$ = 200 GeV. The resulting 
high $p_{T}$ results at mid-rapidity from PHENIX~\cite{phenix_dAu}, STAR~\cite{star_dAu}, 
PHOBOS~\cite{phobos_dAu}, and BRAHMS~\cite{brahms_hipt_200} consistently indicate the following:

\begin{itemize}
\item High $p_{T}$ inclusive $h^\pm$~\cite{phenix_dAu,star_dAu,phobos_dAu,brahms_hipt_200} 
and $\pi^0$~\cite{phenix_dAu} spectra from d+Au minimum bias (MB) collisions
are not suppressed but are {\it enhanced} compared to p+p collisions 
($R_{ dAu}$ plots in Fig.~\ref{fig:dAu}), in a way very much reminiscent of the ``Cronin effect'' 
observed in fixed-target p+A collisions at lower $\sqrt{s}$~\cite{cronin}. 
As a matter of fact, p+Au collisions (from neutron-tagged d+Au events~\cite{phenix_dAu}) 
show a similar behaviour as minimum bias d+Au collisions.
\item Above $p_{T}\sim$ 2.5 GeV/$c$ the nuclear modification factor of inclusive charged
hadrons in MB d+Au collisions saturates at~\footnote{%Actually, 
$R_{dAu}^{_{\mbox{\tiny{PHENIX}}}}(p_{T}=2-7$ GeV/$c)\sim$ 1.35, 
$R_{dAu}^{_{\mbox{\tiny{STAR}}}}(p_{T}=2-6$ GeV/$c)\sim$ 1.45, 
$R_{dAu}^{_{\mbox{\tiny{BRAHMS}}}}(p_{T}=2-5$ GeV/$c)\sim$ 1.3.} $R_{dAu} \sim$ 1.4. 
Above 6 GeV/$c$, STAR $h^\pm$ and PHENIX $\pi^0$ results seem to indicate that $R_{dAu}$ 
decreases as a function of $p_{T}$, becoming consistent with 1 at around 8 GeV/$c$. 
\item The ``Cronin enhancement'' for unidentified hadrons at high $p_{T}$ ($R_{dAu}^{h^\pm}\approx$ 1.35) 
is larger %~\footnote{This has been actually a known experimental observation~\cite{cronin} 
%which has lacked so far a clear theoretical explanation.} 
than for neutral pions ($R_{dAu}^{\pi^0}\approx$ 1.1)~\cite{phenix_dAu}.
\item The degree of ``enhancement'' in d+Au compared to p+p {\it increases} with collision 
centrality~\cite{phobos_dAu,star_dAu}, an opposite trend to Au+Au results. 
\item The azimuthal correlations in MB and central d+Au collisions are very similar to that
of p+p and do not show the significant suppression of the away-side peak observed
in central Au+Au reactions~\cite{star_dAu}.
\end{itemize}

All these results lead to the conclusion that no ``cold'' nuclear matter (or initial-state) 
effects, -like a strong saturation of the nuclear parton distribution functions in the relevant
($x,Q^{2}$) kinematical region probed by the current experimental setups-, can explain
the high $p_{T}$ behaviour in central Au+Au. The data suggest, instead, that
final-state interactions are responsible of the high $p_{T}$ suppression and the disappearance 
of back-to-back dijet correlations observed at mid-rapidity in central Au+Au reactions.

%%%%%%%%%%%%%%%%%%%%%%%%%%%%%%%%%%%%%%%%%%%%%%

\section{Summary}

%MIND that: For pT > 4--5 GeV/$c$: p_{T}, sqrt(s) and x_{T}, h/pi, v2? ~pQCD
A vast body of high transverse momentum data has been collected and analyzed during 
2000--2003 by the 4 experiments at RHIC in Au+Au, p+p, d+Au collisions at 
$\sqrt{s_{_{NN}}}$ = (130)200~GeV. This has permitted a detailed comparative study of the properties 
of high $p_{T}$ particle production in high-energy heavy-ion collisions as a function of 
$p_{T}$, $\sqrt{s_{_{NN}}}$, collision centrality, pseudo-rapidity, and particle species.
All these studies reveal that hadron production at high transverse momentum in central Au+Au 
reactions shows significant deviations compared to p+p, d+Au, and Au+Au 
peripheral collisions at RHIC energies, as well as to nucleus-nucleus data at lower center-of-mass energies. 
The main observations are: (i) the inclusive and identified spectra are suppressed by a factor 
4 -- 5 above $p_{T}\approx$ 5 GeV/$c$ compared to the expectations of $N_{coll}$ scaling, (ii) 
the baryon yields are enhanced compared to the meson yields in the range of intermediate momenta 
$p_{T}\approx$ 2 -- 5 GeV/$c$, (iii) the back-to-back dijet azimuthal correlations 
are significantly suppressed, (iv) there is a strong constant elliptic flow signal at high $p_{T}$.
The whole set of results puts strong experimental constraints on the properties of the underlying 
QCD medium produced in Au+Au reactions at collider energies. Comparison of the energy spectra
and angular correlations data to the theoretical calculations globally supports pQCD-based 
models of final-state parton energy loss in a dense medium, although non-perturbative effects 
(like e.g. quark coalescence) are needed in order to explain the baryon-meson differences in 
yield and $v_{2}$ in the intermediate $p_{T}$ window ($p_{T}\approx$ 2 -- 5 GeV/$c$). Theoretical 
predictions of a strong saturation of the nuclear wave function at high energies are also in 
agreement with most of the data but do not seem to explain consistently Au+Au and d+Au RHIC 
results at midrapidity. Coming ion-ion runs at RHIC and, in the mid-term, Pb+Pb collisions 
at LHC energies will help to further strengthen our current understanding of the physics 
of QCD media at high energy densities.

%%%%%%%%%%%%%%%%%%%%%%%%%%%%%%%%%%%%%%%%%%%%%%

\end{document}